\documentclass{aa}  

\usepackage{graphicx}
\usepackage{txfonts}
\usepackage{colortbl} 
\usepackage[normalem]{ulem} 
\usepackage[]{caption}
\usepackage{subfigure}

\usepackage{multirow}
\newcommand{\Lnom}{\hbox{$\mathcal{L}^{\rm N}_{\odot}$}}
\newcommand{\Mnom}{\hbox{$\mathcal{M}^{\rm N}_{\odot}$}}
\newcommand{\Rnom}{\hbox{$\mathcal{R}^{\rm N}_{\odot}$}}
\begin{document} 

\title{Apsidal motion in the massive binary HD~152\,248}
\author{S.\ Rosu\inst{1}\fnmsep\thanks{Research Fellow FRS-FNRS (Belgium)} \and G.\ Rauw\inst{1} \and K. E.\ Conroy\inst{2}  \and E.\ Gosset\inst{1}\fnmsep\thanks{Research Director FRS-FNRS (Belgium)} \and J.\ Manfroid\inst{1} \and P.\ Royer\inst{3} }
\mail{sophie.rosu@uliege.be}
\institute{Space sciences, Technologies and Astrophysics Research (STAR) Institute, Universit\'e de Li\`ege, All\'ee du 6 Ao\^ut, 19c, B\^at. B5c, 4000 Li\`ege, Belgium
\and
Villanova University, Dept. of Astrophysics and Planetary Sciences, 800 E Lancaster Ave, Villanova PA 19085, USA
  \and
Instituut voor Sterrenkunde, KU Leuven, Celestijnenlaan 200D, Bus 2401, 3001 Leuven, Belgium}
\date{}

  \abstract
   {The eccentric massive binary HD~152\,248 (also known as V1007~Sco), which hosts two O7.5~III-II(f) stars, is the most emblematic eclipsing O-star binary in the very young and rich open cluster NGC~6231. Its properties render the system an interesting target for studying tidally induced apsidal motion.}{Measuring the rate of apsidal motion in such a binary system gives insight into the internal structure and evolutionary state of the stars composing it.}{A large set of optical spectra was used to reconstruct the spectra of the individual binary components and establish their radial velocities using a disentangling code. Radial velocities measured over seven decades were used to establish the rate of apsidal motion. We furthermore analysed the reconstructed spectra with the {\tt CMFGEN} model atmosphere code to determine stellar and wind properties of the system. Optical photometry was analysed with the {\tt Nightfall} binary star code. A complete photometric and radial velocity model was constructed in {\tt PHOEBE~2} to determine robust uncertainties.}{We find a rate of apsidal motion of $(1.843^{+0.064}_{-0.083})^{\circ}$\,yr$^{-1}$. The photometric data indicate an orbital inclination of $(67.6^{+0.2}_{-0.1})^\circ$ and Roche-lobe filling factors of both stars of about $0.86$. Absolute masses of $29.5^{+0.5}_{-0.4}~\text{M}_\odot$  and mean stellar radii of $15.07^{+0.08}_{-0.12}~\text{R}_\odot$ are derived for both stars. We infer an observational value for the internal structure constant of both stars of $0.0010\pm 0.0001$.} {Our in-depth analysis of the massive binary HD~152\,248 and the redetermination of its fundamental parameters can serve as a basis for the construction of stellar evolution models to determine theoretical rates of apsidal motion to be compared with the observational one. In addition, the system hosts two twin stars, which offers a unique opportunity to obtain direct insight into the internal structure of the stars.}
\maketitle
%

\section{Introduction}
The very young and rich open cluster NGC~6231 \citep[][and references therein]{Sung,Reipurth,Kuhn} hosts a population of 15 massive O-type objects with a high incidence of short-period binaries \citep{Sana08}. The most emblematic O-star binary in NGC~6231 is probably the O7.5\,III(f) + O7\,III(f) system HD~152\,248 \citep{Sana01,Mayer08}. The binarity of this star was discovered by \citet{Struve}, and its orbital period near 6\,d was first established by \citet{Hill}. Subsequent studies refined the value of the orbital period to close to 5.816\,d \citep{Mayer92,Stickland,Sana01,Nesslinger06,Mayer08}. The existence of photometric eclipses in the light curve of HD~152\,248 was first suspected by \citet{Brownlee} and subsequently confirmed by \citet{Mayer92}. Various studies concluded that the orbit has an eccentricity of $e \sim 0.10$ -- $0.15$ \citep{Mayer92,Stickland,Penny,Sana01,Mayer08}. Stellar wind interactions have been found both in optical spectroscopy \citep{Sana01} and through phase-locked variations of the X-ray emission \citep{Sana04}. HD~152\,248 is a remarkable example of a twin system with a short orbital period and a mass ratio very close to unity \citep{Sana01,Mayer08}. Such systems are important diagnostics in the context of 
the theory of star formation (e.g., \citet{tokovinin}). They could be formed through orbital decay of wider binary systems as a result of gas accretion or dynamical interaction with a circumbinary disk \citep{bate}.

Interestingly, some previous studies reported evidence of a secular variation of the argument of periastron $\omega$ \citep{Sana01,Nesslinger06,Mayer08}. These properties make HD~152\,248 an ideal target for the study of apsidal motion \citep{Schmitt,Rauw,Zasche}. Apsidal motion consists in a slow precession of the line of apsides in an eccentric binary and stems from the non-spherical shape of the gravitational field of stars in close binaries \citep[][and references therein]{Schmitt}. The rate of apsidal motion $\dot{\omega}$ depends upon the internal structure of the stars that make up the binary system \citep[e.g.,][]{Shakura,CG10}. Measuring the rate of apsidal motion thus provides valuable information about the internal structure of stars. This is especially important for the most massive stars, for which such information is rather scarce \citep{Bulut}. 

We here perform a detailed study of the apsidal motion of HD~152\,248. In accordance with previous studies \citep[e.g.,][]{Stickland,Penny,Sana01,Mayer08}, we call the primary star the one that is eclipsed during primary minimum. The extensive set of observational data that we use is introduced in Sect.\,\ref{observations}. In Sect.\,\ref{specanalysis} we perform the spectral disentangling and analyse the reconstructed spectra of the binary components by means of the {\tt CMFGEN} model atmosphere code \citep{Hillier}.  The radial velocities inferred from the spectral disentangling are combined with data from the literature in Sect.\,\ref{omegadot} to establish a preliminary value for the observational rate of apsidal motion. The possibility of a triple system is discussed in Appendix\,\ref{appendix}. Photometric data of HD~152\,248 are analysed in Sect.\,\ref{photom} by means of the {\tt Nightfall} binary star code. In Sect.\,\ref{sect:phoebe} we construct a complete model  in PHOEBE~2 that takes photometric and radial velocity (RV) measurements into account. We determine robust uncertainties on the parameters based on this model. The observational internal structure constant $k_2$ of the stars is determined in Sect.\,\ref{sect:k2}. Finally, we discuss the implications of our study and give our conclusions in Sect.\,\ref{conclusion}.

\section{Observational data \label{observations}}
\subsection{Spectroscopy}
To investigate the optical spectrum of HD~152\,248, we extracted 48 high-resolution \'echelle spectra. Of these, 27 were obtained with the Fiber-fed Extended Range Optical Spectrograph (FEROS) mounted on the European Southern Observatory (ESO) 1.5\,m or 2.2\,m telescopes in La Silla, Chile \citep{kaufer}. The remaining 21 spectra were obtained with the ESPaDOns spectrograph attached to the Canada-France-Hawa\"i observatory (CFH) 3.6\,m telescope in Hawa\"i \citep{Donati}. These data were collected between May 1999 and April 2014. The FEROS instrument has a spectral resolving power of 48\,000, and its detector is an EEV CCD with 2048\,$\times$\,4096 pixels of 15\,$\times$\,15\,$\mu$m. Thirty-seven orders cover the wavelength domain from 3650 to 9200\,\AA. Exposure times range from 150 to 700 seconds. The ESPaDOns instrument has a spectral resolving power of 68\,000, and its detector is a CCD with 2k\,$\times$\,4.5k pixels of 13.5\,$\times$\,13.5\,$\mu$m. Forty orders cover the wavelength domain from 3700 to 10\,500\,\AA~(with three small gaps: 9224-9234\,\AA, 9608-9636\,\AA, and 10\,026-10\,074\,\AA). Exposures times were 540 seconds. The FEROS data were reduced using the FEROS pipeline of the {\tt MIDAS} software. The reduced ESPaDOnS data were retrieved from the CFHT archive. Residual cosmic rays were removed within {\tt MIDAS}, and the {\it telluric} tool within {\tt IRAF} was used along with the atlas of telluric lines of \citet{Hinkle} to remove the telluric absorptions. The spectra were normalised with {\tt MIDAS} by fitting low-order polynomials to the continuum. The journal of the spectroscopic observations is presented in Appendix\,\ref{appendix:spectrotable}, Table\,\ref{Table:spectro+RV}.

\subsection{Photometry}
\label{sect:photo}
HD~152\,248 was observed between 22 March and 19 April 1997 with the 0.6\,m Bochum telescope at La Silla observatory, Chile. The Cassegrain focus of the telescope was equipped with an imaging camera featuring a CCD detector with a field of view of 3.2 arcmin\,$\times$\,4.8\,arcmin. We used two narrow-band filters, designed for the study of Wolf-Rayet stars \citep{Royer}. The central wavelengths and full widths at half-maximum (FWHM) of these filters were 4684\,\AA\ and 30\,\AA\ for the He\,{\sc ii} $\lambda$\,4686 filter, and 6051\,\AA\ and 28\,\AA\ for the $c_2$ continuum filter. Dome flats were collected each day, and bias exposures were taken at various times over the night. An extensive description of the data reduction can be found in \citet{Sana05}. The rms of the differences between the comparison stars ranged between 0.008 and 0.010\,mag, and we therefore adopted 0.010\,mag as an estimate of the measurement error. Our dataset consists of 125 data measurements in the $c_2$ filter and 103 measurements in the He\,{\sc ii} $\lambda$\,4686 filter. The corresponding data are given in Appendix\,\ref{appendixB}, Tables\,\ref{journalHeII} and \ref{journalc2}.

Most recently, HD~152\,248 was observed by NASA's Transiting Exoplanet Survey Satellite \citep[{\it TESS},][]{TESS}. The instrument on board {\it TESS} has a bandpass ranging from 6000\,\AA\, to 1\,$\mu$m \citep{TESS}. The 15\,$\times$\,15\,$\mu$m pixels of the CCD detectors correspond to 21\arcsec~on the sky, thereby undersampling the instrument point spread function (PSF). Stars as bright as our target saturate the central pixel, but the excess charges created are spread into adjacent pixels through the blooming effect, allowing us in principle to recover the photometry for stars up to fourth magnitude. The {\it TESS} pipeline \citep{Jen16} is based on the {\it Kepler} pipeline \citep{Jen10}. Simple background-corrected aperture photometry was extracted and subsequently corrected for by removing signatures in the light curve that correlate with systematic spacecraft or instrument effects. We extracted the {\it TESS} light curve of HD~152\,248 corrected in this way and discarded any data points that had a quality flag different from 0. This resulted in a time series of 12\,372 measurements with a nominal time step of 2 minutes and ranging over two time intervals from HJD\,2\,458\,629.86 to 2\,458\,639.00 (6372 data points) and from HJD\,2\,458\,644.44 to 2\,458\,652.90 (6000 data points). Most probably as a result of the severe crowding in the NGC~6231 cluster and the complicated pattern produced by saturated targets, the pipeline-processed {\it TESS} light curve displays a number of artifacts that have amplitudes well beyond the theoretical photometric accuracy (which should be about 0.0002\,mag).

\section{Spectral analysis\label{specanalysis}}
This section is devoted to the spectral analysis of the binary system. Spectral disentangling was performed in order to reconstruct the individual spectra and determine the radial velocities of the stars over the orbital cycle. The spectral classification was reassessed and the projected rotational velocities were derived for both stars. The reconstructed spectra were then fit with non-local thermal equilibrium (non-LTE) synthetic {\tt CMFGEN} spectra in order to constrain the fundamental properties of the stars (effective temperatures, surface gravities, chemical compositions, and mass-loss rates).
\subsection{Spectral disentangling}
\subsubsection{Method}
To reconstruct the spectra of the individual components of HD~152\,248 and simultaneously establish their radial velocities (RVs) as a function of time, we used our spectral disentangling code, which is based on the method described by \citet{GL}. The individual spectra were reconstructed in an iterative manner. We consider that we work on the spectrum of star A. At each step of the iterative procedure, we first subtracted the current best approximation of the reconstructed spectrum of star B shifted to the current estimate of $RV_\text{B}(t)$ from the observed spectrum at time $t$. The residual spectra built in this way were then shifted into the frame of reference of star A and averaged to obtain a new estimate of the spectrum of star A. Updated estimates of $RV_\text{A}(t)$ were obtained by cross-correlating the residual spectra with a synthetic {\tt TLUSTY} spectrum \citep{LH} and fitting a parabola to the peak of the correlation function to determine its centre \citep{VD}. For the synthetic {\tt TLUSTY} spectra, we assumed effective temperatures of 35\,000~K, $\log{g}=3.5,$ and $v\,\sin{i} = 150$\,km\,s$^{-1}$ for both stars. The whole procedure was then repeated inverting the roles of stars A and B. To start the iterations, we first assumed a featureless spectrum for star B with a flux level of half the normalised flux of the combined spectrum. To check the robustness of the results of the disentangling against any bias that the initial approximation of the spectrum of star B could introduce, we repeated the disentangling procedure by interchanging the roles of stars A and B. The agreement was very good and the final reconstructed spectra were taken to be the mean of the two approaches. 

To avoid any biases of the reconstruction process due to the eclipses, we restricted ourselves to those spectra (40 out of the 48) for which the raw data indicate a significant separation of the components, that is,\ well away from conjunction phases. We performed a total of 60 iterations to disentangle the spectra and determine the RVs. Because the two stars have nearly identical spectra, the disentangling code might become confused and accidentally assign the primary RV to the secondary and vice versa. To avoid this problem and increase the robustness of the code, the first 30 iterations were performed with the RVs fixed to the initial input values estimated using a Gaussian fitting of the He\,{\sc i} $\lambda\lambda$\,4026, 4471, He\,{\sc ii} $\lambda\lambda$\,4200, 4542, 5412, and O\,{\sc iii} $\lambda$\,5592 lines. In this manner, the individual spectra were mostly reconstructed after 30 iterations. The last 30 iterations were then performed as explained in the previous paragraph.  

The spectral disentangling was performed separately over a number of wavelength domains: A0[3800:3920]\,\AA, A1[3990:4400]\,\AA, A2[4300:4570]\,\AA, A3[4570:5040]\,\AA, A4[5380:5760]\,\AA, A5[5380:5860]\,\AA, A6[5830:6000]\,\AA, A7[6400:6750]\,\AA,~and A8[7000:7100]\,\AA. The presence of interstellar lines or diffuse interstellar bands close to spectral lines in some of these spectral domains (A0, A5, A6, and A7) affects the quality of the resulting reconstructed spectra.
In these cases, the disentangling code erroneously associated some of the line flux of the non-moving interstellar medium (ISM) lines to the stars or vice versa. This situation can also affect the determination of the stellar RVs. Therefore, we proceeded as follows: we first processed the wavelength domains (A1, A2, A3, A4, and A8) for which the code was able to reproduce the individual spectra and simultaneously estimate the RVs of the stars. We then computed a weighted-average mean of the stellar RVs. The RVs from the individual wavelength domains were weighted according to the number of strong lines present in these domains (five lines for A1, three for A2, one for A3, three for A4, and one for A8). The resulting RVs of both stars are reported in Table\,\ref{Table:spectro+RV} together with their $1\sigma$ errors. We finally performed the disentangling on the remaining four domains (A0, A5, A6, and A7) with RVs fixed to these weighted averages and using a version of the code designed to deal with a third spectral component \citep{Mahy12}. 

\subsubsection{Limitations of the method \label{limitdisent}}
Before we turn to the {\tt CMFGEN} analysis (see Sect.\,\ref{fitcmfgen}), it is important to recall some characteristics of the disentangling method that might affect the forthcoming analysis. Spectral disentangling can only reconstruct features that are well sampled by the Doppler excursions during the orbital cycle. In practice, this often leads to artifacts in the wings of broad lines, such as the H\,{\sc i} Balmer lines, if these lines only partially deblend over the orbital cycle \citep[e.g.,][]{Raucq}. However, in the case of HD~152\,248, the RV amplitudes are sufficient to ensure a reliable reconstruction of the wings of the Balmer absorption lines. 

Stars in a close binary are not spherically symmetric, but are better described by the shape of the equipotentials of the Roche potential $\Omega$. The surface gravity is then given by $|\vec{\nabla}(\Omega)|,$ implying a complex surface gravity distribution. Gravity darkening therefore leads to a non-uniform temperature distribution \citep{Palate12,Palate13}. As the stars revolve around each other, the non-uniform temperature distribution induces a phase-dependence of the strengths of the spectral lines seen by an external observer. Moreover, in an eccentric binary such as HD~152\,248, the deviation of the stellar shape from spherical symmetry changes as a function of orbital phase, and this leads to a dynamical deformation of the stellar surface that can induce line profile variations \citep{Palate13}, see also Figs.\,2, 3, and 4 from \citet{Sana01}. The spectra obtained by disentangling provide an average over the parts of the stellar surface that are seen by the observer and over time. 

Finally, some of the emission lines (He\,{\sc ii} $\lambda$\,4686 and H$\alpha$) seen in the spectrum of HD~152\,248 are at least partially formed in the wind interaction zone of the system \citep{Sana01}. When a spectral disentangling method is applied, we implicitly assume that each spectral feature belongs to one of the stars. For these emission lines, this leads to line shapes and fluxes in the reconstructed spectra that are not those of the stars themselves. In the spectral analysis, these lines are commonly used as diagnostics of the stellar wind properties \citep[mass-loss rates, clumping, wind velocity, etc., see, e.g.,][and references therein]{Martins11}. In the present case, we can only use them as an indication: if the level of emission predicted by the model exceeds the strength of the lines in the reconstructed spectrum, then the mass-loss rate of the model is likely too high. 

\subsection{Spectral classification and absolute magnitudes \label{spectraltype}}
The reconstructed spectra of the binary components allowed us to reassess the spectral classification of the stars. We used Conti's criterion \citep{Conti71} complemented by \citet{Mathys88} based on the ratio between the equivalent widths (EWs) of the He\,{\sc i} $\lambda$\,4471 and He\,{\sc ii} $\lambda$\,4542 lines to determine the spectral type of the stars. We found that $\log W' = \log[\text{EW(He\,{\sc i} $\lambda$\,4471)/EW(He\,{\sc ii} $\lambda$\,4542)}]$ amounts to $0.07\pm 0.01$ and $0.09\pm 0.01$ for the primary and secondary stars, respectively, which corresponds to spectral type O7.5 for both stars. Hereafter, errors are given as $\pm1\sigma$. To assess the luminosity class of the stars, we used the Walborn and Sota criteria \citep{WP90,Sota11,Sota} based on the strengths of the He\,{\sc ii} $\lambda$\,4686 and N\,{\sc iii} $\lambda\lambda$\,4634-40-42 lines. Because the N\,{\sc iii} triplet is in weak emission and the He\,{\sc ii} line neither in strong absorption nor strong emission, the luminosity class is between III and II. This is confirmed by the intensity of the He\,{\sc i} $\lambda$\,4471 line, which is lower than the intensity of the H\,$\gamma$ line. This excludes luminosity class I. As a result, we determine that both stars are of O7.5~III -- II(f) type. 

The brightness ratio in the visible domain was estimated based on the ratio between the EWs of the spectral lines of the primary and secondary stars. For this purpose, we used the H$\gamma$, He\,{\sc i} $\lambda\lambda$\,4026, 4471 and He\,{\sc ii} $\lambda\lambda$\,4200, 4542 lines. The value obtained is $l_1/l_2 = 1.000 \pm 0.025$.

The second {\it Gaia} data release \citep[DR2,][]{Brown} quotes a parallax of $\varpi = 0.59 \pm 0.06$\,mas, corresponding to a distance of $1631^{+178}_{-147}$\,pc \citep{bai18}. From this distance, we derive a distance modulus of $11.06^{+0.24}_{-0.20}$. \citet{hog} reported mean $V$ and $B$ magnitudes of $6.05\pm0.01$ and $6.15\pm0.01,$ respectively. Adopting a value of $-0.27\pm0.01$ for the intrinsic colour index $(B-V)_0$ of an O7.5 III star \citep{MartinsPlez} and assuming the reddening factor in the V-band $R_V$ equal to $3.3\pm0.1$ \citep{Sung}, we infer an absolute magnitude in the V-band of the binary system $M_V=-6.23^{+0.25}_{-0.21}$. The brightness ratio then yields individual absolute magnitudes $M_{V,1}=M_{V,2} =-5.48^{+0.25}_{-0.21}$ for both stars. Comparing these magnitudes to those reported by \citet{Martins05} and \citet{MartinsPlez} shows that the supergiant class I is clearly excluded. The absolute magnitudes are closer to those of an O8 III star rather than a O7.5 III star, although the $M_V$ of an O7.5~III star lies within the error bars.

As pointed out in Sect.\,\ref{limitdisent}, the acceleration of gravity at the surface of the components of a close binary is not given by $\frac{G\,m_*}{R_*^2}$, but instead by $|\vec{\nabla}(\Omega)|$. Therefore, the spectrum of a star in a close binary might mimic that of an object with higher luminosity \citep{Palate12}. In the case of HD~152\,248, this probably explains the difference between the giant classification obtained from photometric analyses and the luminosity class III -- II obtained through spectral classification criteria. 

\subsection{Projected rotational velocities\label{sect:vsini}}
The projected rotational velocities of both stars were derived using the Fourier transform method \citep{Simon-Diaz,Gray}, which has the advantage of being able to separate the effect of rotation from other broadening mechanisms such as macroturbulence. We applied this method to seven well-isolated spectral lines, which are therefore expected to be free of blends. The results are presented in Table\,\ref{vsiniTable}, and the Fourier transform are illustrated for the seven lines in Appendix\,\ref{appendix:fourier}, Fig.\,\ref{Fig:vsini}\subref{subfig:SiIV}-\subref{subfig:HeI5016}. It is commonly recommended to use metallic lines to compute the projected rotational velocity. The H\,{\sc i}, He\,{\sc i} and He\,{\sc ii} line profiles are indeed more affected by non-rotational broadening mechanisms such as Stark broadening, which alter the position of the first zero in the Fourier transform~\citep{Simon-Diaz}. However, when the projected rotational velocity of the star exceeds 100 km\,s$^{-1}$, the effect of Stark broadening on the He\,{\sc i} lines is expected to be small and these lines can then be used as a complement to the metallic lines. The results presented in Table\,\ref{vsiniTable} show that the mean $v \sin i_\text{rot}$ computed on metallic lines alone or on all the lines agree very well. We adopt a mean $v \sin i_\text{rot}$ of $138\pm4$ km\,s$^{-1}$ for the primary star and $137\pm6$ km\,s$^{-1}$ for the secondary star. These numbers agree with the values obtained by \citet{Mayer08}, but are lower than those obtained by the other authors (see Table\,\ref{vsiniTable}). This result is expected because the cross-correlation method these authors used does not separate the rotational broadening from other broadening mechanisms. \\

\begin{table}[h!]
\caption{Best-fit projected rotational velocities as derived from the disentangled spectra of HD~152\,248.}
\begin{center}
\begin{tabular}{l l l}
\hline\hline
Line & \multicolumn{2}{c}{$v\sin i_{\text{rot}}$ (km\,s$^{-1}$)} \\
& Primary & Secondary \\
\hline
Si\,{\sc iv} 4089 & 132 & 138 \\ 
O\,{\sc iii} 5592 & 142 & 143 \\ 
C\,{\sc iv} 5812 & 140 & 129 \\ 
He\,{\sc i} 4120 & 130 & 133 \\ 
He\,{\sc i} 4713 & 153 & 153 \\ 
He\,{\sc i} 4922 & 128 & 115 \\ 
He\,{\sc i} 5016 & 141 & 142 \\
\hline
Mean (metallic lines) & $138\pm4$ &$137\pm6$ \\ 
Mean (all lines) & $138\pm8$ & $136\pm11$ \\
\hline
\citet{Stickland} & $190\pm20$ & $190\pm20$ \\
\citet{Howarth} & $159\pm14$ & $165\pm14$ \\
\citet{Penny} & $163$ & $173$ \\
\citet{Mayer08} & $135\pm5$ & $135\pm5$ \\
\hline
\end{tabular}
\end{center}
\tablefoot{The values quoted by \citet{Stickland}, \citet{Howarth}, and \citet{Penny} were obtained by cross-correlation techniques applied to \textit{IUE} spectra. The values of \citet{Mayer08} were obtained through comparison of the He\,{\sc i} $\lambda$\,4922 line with rotationally broadened synthetic spectral lines.}  
\label{vsiniTable}
\end{table}

\subsection{Model atmosphere fitting \label{fitcmfgen}}
The reconstructed spectra of the binary components were analysed by means of the {\tt CMFGEN} model atmosphere code \citep{Hillier} to constrain the fundamental properties of the stars. This code solves the radiative transfer, that is, statistical and radiative equilibrium equations in the comoving frame assuming the star and its wind are spherically symmetric. Line-blanketing and clumping are included in the code. For the wind, a standard $\beta$-velocity law is adopted through the relation
\begin{equation}
v(r) = v_\infty\left(1-\frac{R_*}{r}\right)^\beta,
\end{equation}
where $v_\infty$ is the terminal wind velocity and $R_*$ the radius of the star. The wind clumping is introduced in the models using the two-parameter exponential law for the volume filling factor
\begin{equation}
f(r) = f_1+(1-f_1) \exp\left(-\frac{v(r)}{f_2}\right),
\end{equation}
where the parameters $f_1$ and $f_2$ are the filling factor value when $r \rightarrow \infty$ and the onset velocity of clumping, respectively. Including clumping in the models leads to a reduced estimate of the mass-loss rate compared to an unclumped model. To compare with values obtained with unclumped models, the mass-loss rate derived in this paper has to be multiplied by a factor of $1/\sqrt{f_1}$~\citep{Martins11}. Regarding the microturbulent velocity $v_\text{micro}$, {\tt CMFGEN} assumes that it depends on the position $r$ as 
\begin{equation}
v_\text{micro} = v_\text{micro}^\text{min} + \left(v_\text{micro}^\text{max} - v_\text{micro}^\text{min}\right) \frac{v(r)}{v_\infty},
\end{equation}
where $v_\text{micro}^\text{min}$ and $v_\text{micro}^\text{max}$ are the minimum and maximum $v_\text{micro}$ whose values are fixed to 15~km\,s$^{-1}$ and $0.1v_\infty$.  \\

The {\tt CMFGEN} spectra were first broadened by the projected rotational velocities determined in Sect.\,\ref{sect:vsini}. The macroturbulence velocity was then adjusted on the wings of the O\,{\sc iii} $\lambda$\,5592 line. The stellar parameters (effective temperature, surface gravity, and surface abundances) as well as the wind parameters (mass-loss rate, wind terminal velocity, and clumping factors) were then adjusted following the procedure outlined by \citet{Martins11}.  \\

The effective temperature was determined by searching for the best fit of the He\,{\sc i} $\lambda\lambda$\,4026, 4471, 4713, 4920, 5016, 5876, and 7065 and He\,{\sc ii} $\lambda\lambda$\,4200, 4542, and 5412 lines. The resulting effective temperatures are $34\,000 \pm 1000$ K for both stars. 

The surface gravity was obtained by adjusting the wings of the Balmer lines H$\beta$, H$\gamma$, H$\delta,$ and H\,{\sc i} $\lambda\lambda$\,3835, 3890. We excluded the Balmer line H$\varepsilon$ because this line is not correctly reproduced by the disentangling code because of the blend with the interstellar Ca\,{\sc ii} line. We obtain $\log g_\text{spectro} = 3.48\pm0.10$ for both stars.  \\

\begin{figure*}[htbp]
\begin{center}
\includegraphics[width=\linewidth]{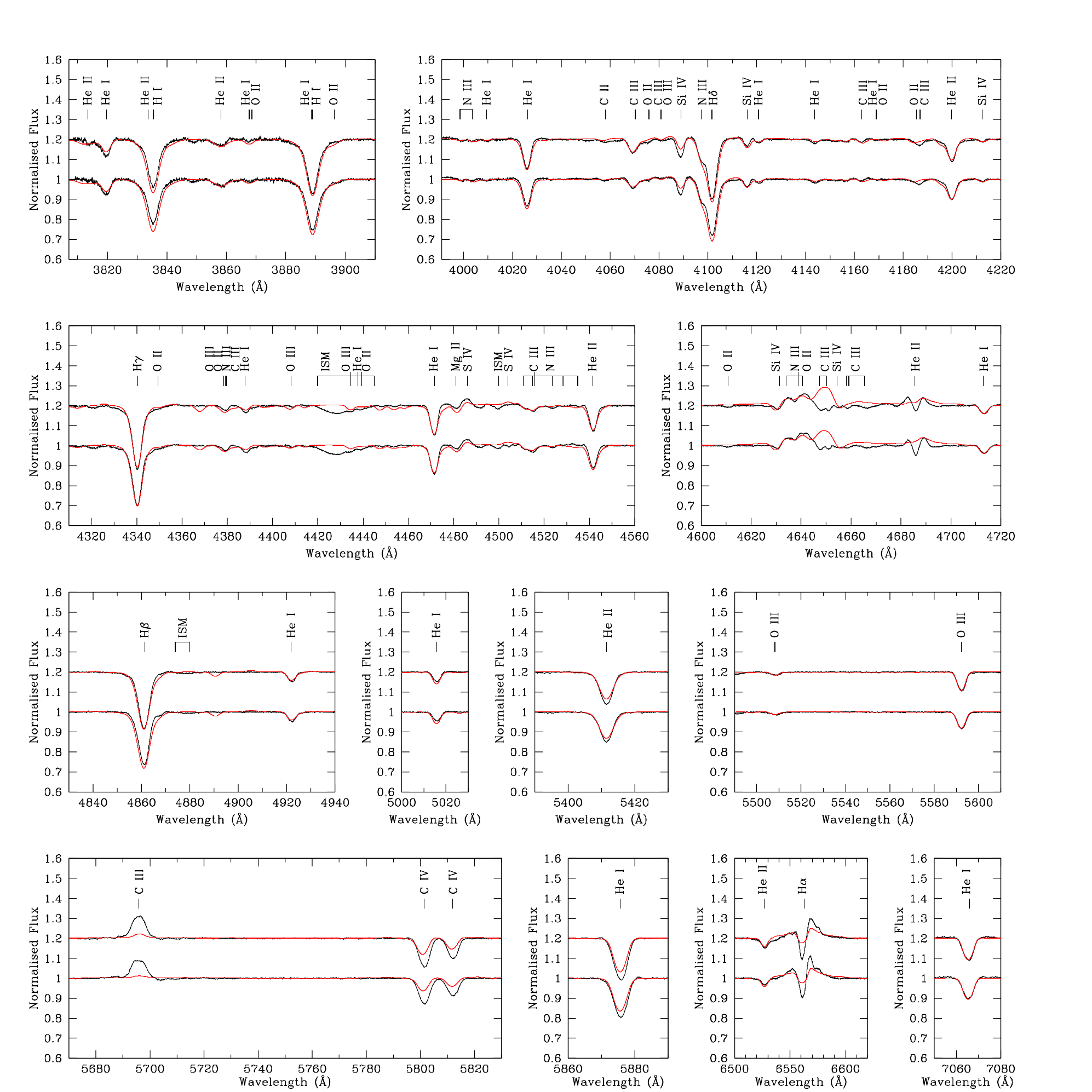}
\end{center}
\caption{Normalised disentangled spectra (in black) of the primary and secondary star of HD~152\,248 (note that the spectrum of the secondary star is shifted by +0.2 in the $y$-axis for convenience) together with the respective best-fit {\tt CMFGEN} model atmosphere (in red). \label{Fig:CMFGEN}}
\end{figure*}

The surface chemical abundances (in number compared to H) of C, N, and O were adjusted on corresponding lines, but we decided to set the abundances of the other elements, including He, to solar abundances as taken from \citet{Asplund}.\\ 
For the CNO elements, a $\chi^2$ analysis was performed to obtain an estimate of their abundances and associated uncertainties. We used the method proposed by \citet{Martins15}: for each of the three elements, we ran four models in addition to the one taken as a reference. For each {\tt CMFGEN} model $m$ and each diagnostic line $l$, we computed the $\chi^2$ as follows:
\begin{equation}
\chi^2_{l,m} = \sum_{\lambda=\lambda_{\text{min},l}}^{\lambda_{\text{max},l}} \left(F_\text{obs}(\lambda)-F_{\text{\tt CMFGEN},m}(\lambda)\right)^2,
\end{equation}
where $F_\text{obs}$ is the normalised observed flux of the star and $F_\text{\tt CMFGEN}$ is the normalised flux of the {\tt CMFGEN} model. For each line, we determined the minimum of the $\chi^2_l$ function by fitting a cubic spline to the values of the five {\tt CMFGEN} models. We renormalised this function by its minimum so that $\chi^2_{l,\text{min}}$ is now equal to 1.  Following \citet{Martins15}, the $1\sigma$ uncertainty on the abundances was then obtained from the abundance values corresponding to $\chi^2=2$ (but see the caveats of \citet{Andrae}).     \\
The oxygen abundance was determined by adjusting the O\,{\sc iii} $\lambda$\,5592 line and checking it with the O\,{\sc iii} $\lambda$\,5508 line. 
The best fit is obtained for an abundance O/H of $4.29^{+1.07}_{-0.70}\times10^{-4}$ and $6.01^{+1.81}_{-1.21}\times10^{-4}$ for the primary and secondary star, respectively. Within the error bars, both stars have a solar oxygen abundance. The synthetic {\tt CMFGEN} spectra display a series of O\,{\sc iii} absorption lines ($\lambda\lambda$\,4368, 4396, 4448, 4454, and 4458) that are not present in the observed spectra (neither before nor after disentangling). These lines are not correctly reproduced by the model, as has been pointed out by several authors \citep[e.g.,][]{Raucq} and were not considered for the determination of the oxygen abundance.\\
The nitrogen abundance was determined by adjusting the N\,{\sc iii} $\lambda\lambda$\,4510-4530 lines. Special attention was paid to the fact that the N\,{\sc iii} $\lambda\lambda$\,4510-4530 complex is polluted by the C\,{\sc iii} $\lambda$\,4516 line. We did not use the N\,{\sc iii} $\lambda\lambda$\,4634-4640 lines in this exercise because they are blended with Si\,{\sc iv}, O\,{\sc ii,} and C\,{\sc iii} lines. In this way, we determined a nitrogen abundance N/H of $1.32^{+0.90}_{-0.72}\times10^{-4}$ and $1.15^{+0.69}_{-0.55}\times10^{-4}$ for the primary and secondary star, respectively. Both stars hence display a nitrogen enrichment compared to the solar abundance. This situation is quite common for somewhat evolved O stars and likely stems from the various internal mixing processes that exist in the interiors of these stars~\citep[][and references therein]{Song}.  
Finally, the carbon abundance was adjusted on the C\,{\sc iii} $\lambda$\,4070 line. We obtain a depleted carbon abundance of $1.17^{+0.30}_{-0.27}\times10^{-4}$ for the primary star and of $2.12^{+0.31}_{-0.28}\times10^{-4}$ for the secondary star. We note that the C\,{\sc iii} $\lambda\lambda$\,4647-51 blend and the C\,{\sc iii} $\lambda$\,5696 emission are not well reproduced by our model. These lines are known to be problematic because their formation processes are complex and are controlled by a number of other far-UV lines \citep{Martins12}. As a result, the strength and nature of C\,{\sc iii} $\lambda\lambda$\,4647-51 and C\,{\sc iii} $\lambda$\,5696 critically depend on subtle details of the stellar atmosphere model. We therefore decided not to consider these lines in our fitting procedure. We further note that the C\,{\sc iv} $\lambda$\,5801 and $\lambda$\,5812 lines are not well reproduced either by our model. These two lines are usually problematic and rarely correctly reproduced \citep[F.\ Martins 2019, private communication, see also][]{Raucq17}. Therefore, we did not use these lines to adjust the carbon abundance. These chemical compositions suggest that the primary star is the slightly more evolved component of the system.   \\

Regarding the wind parameters, the clumping parameters were fixed: the volume filling factor $f_1$ was set to 0.1, and the $f_2$ parameter that determines the wind velocity (and thus the position) where clumping starts was set to 100~km\,s$^{-1}$. For the sake of completeness, we varied $f_1$ from 0.09 to 0.11 and $f_2$ from 70 to 130~km\,s$^{-1}$. We observed that decreasing (increasing) $f_1$ led to a small increase (decrease) of the H\,$\alpha$ emission while decreasing (increasing) $f_2$ led to a small increase (decrease) of both H\,$\alpha$ and He\,{\sc ii} $\lambda$\,4686 emission lines. The exact values of these two parameters do not affect the results of our study because the two affected lines are not used to constrain the parameters of the stars (except for the mass-loss rate, see below).
Likewise, the $\beta$ parameter of the velocity law was fixed to the value of 1.07. This value is between values of O7.5 giant and supergiant stars as inferred by \citet{Muijres}. Again for the sake of completeness, we varied this parameter from 1.00 to 1.15. We observed that decreasing (increasing) $\beta$ led to a small increase (decrease) in H\,$\beta$ and a small decrease (increase) in H\,$\alpha$ and He\,{\sc ii} $\lambda$\,4686 emission lines. When all other parameters were fixed, the best adjustement of H\,$\beta$ was achieved for $\beta=1.07$, suggesting that the adopted $\beta$-value is representative of the system or is at least not too far from the exact value.

In principle, the wind terminal velocity could be derived from the H$\alpha$ line. However, in the spectra of HD~152\,248, this line is at least partially formed in the wind collision zone. Therefore we decided to fix the value of $v_\infty$ for each star to 2420\,km\,s$^{-1}$ as determined by \citet{Howarth} on combined {\it IUE} spectra. Regarding the mass-loss rate, the main diagnostic lines in the optical domain are H$\alpha$ and He\,{\sc ii} $\lambda$\,4686. However, as for H$\alpha$, the weak He\,{\sc ii} emission in the spectra of HD~152\,248 is again at least partially formed in the wind collision zone, leading to an enhanced emission strength. To obtain a first-order estimate, we thus decided to adjust the mass-loss rate on the depth of the H$\gamma$ and H$\delta$ lines. As a second step, we lowered the mass-loss rate in order to avoid H$\alpha$ and He\,{\sc ii} $\lambda$\,4686 emissions in the synthetic spectrum that are stronger than in the disentangled spectra. In this way, we found that $\dot{\text{M}} \le 8\times10^{-7}~\text{M}_\odot/\text{yr}$ for both stars. \\

The stellar and wind parameters of the best-fit {\tt CMFGEN} model atmosphere are summarised in Table\,\ref{Table:CMFGEN} and the chemical abundances are given in Table\,\ref{Table:CMFGEN_ab}. The normalised disentangled spectra of the binary components of HD~152\,248 are illustrated in Fig.\,\ref{Fig:CMFGEN} along with the best-fit {\tt CMFGEN} adjustment.    \\

\begin{table}[h!]
\caption{Stellar and wind parameters of the best-fit {\tt CMFGEN} model atmosphere derived from the separated spectra of HD~152\,248.}
\begin{center}
\begin{tabular}{l c c}
\hline\hline
Parameter & \multicolumn{2}{c}{Value} \\
& Primary & Secondary \\
\hline
$\text{T}_\text{eff}$ (K) & $34\,000\pm1000$ & $34\,000\pm1000$  \\ 
$\log g_\text{spectro}$ (cgs) & $3.48\pm0.10$ & $3.48\pm0.10$ \\
$v_\text{macro}~(\text{km\,s}^{-1})$ & $130\pm10$ & $120\pm10$ \\ 
$\dot{\text{M}}~(\text{M}_\odot\,\text{yr}^{-1})$  & $\le 8\times10^{-7}$ &$\le 8\times10^{-7}$ \\
$\dot{\text{M}}_\text{unclumped}~(\text{M}_\odot\,\text{yr}^{-1})^a$  & $\le 2.5\times10^{-6}$ & $\le 2.5\times10^{-6}$ \\
$v_\infty~(\text{km\,s}^{-1})$ & $2420^b$  & $2420^b$ \\
$f_1$ & 0.1 (fixed) & 0.1 (fixed)  \\ 
$f_2~(\text{km\,s}^{-1})$ & 100 (fixed) & 100 (fixed)\\
$\beta$ & 1.07 (fixed) & 1.07 (fixed) \\
\vspace*{-4mm}\\
$R_\text{spectro}\,(\text{R}_{\odot})$ & $13.7^{+1.9}_{-1.6}$ & $13.7^{+1.9}_{-1.6}$ \\
\vspace*{-3mm}\\
$\text{M}_\text{spectro}\,(\text{M}_\odot)$ & $20.8^{+7.4}_{-6.9}$ & $20.8^{+7.4}_{-6.9}$ \\
\vspace*{-3mm}\\
\hline
\end{tabular}
\end{center}
\tablefoot{ $^a$$\dot{\text{M}}_\text{unclumped}=\dot{\text{M}}/\sqrt{f_1}$. $^b$Value fixed to the value determined by \citet{Howarth}.}
\label{Table:CMFGEN}
\end{table} 

\begin{table*}[htbp]
\caption{Surface chemical abundances of the best-fit {\tt CMFGEN} model atmosphere derived from the separated spectra of HD~152\,248 (Cols.\,2 and 3). The solar chemical abundances are given in Col.\,4 for comparison and are taken from \citet{Asplund}. }
\begin{center}
\begin{tabular}{l c c c}
\hline\hline
Element & Primary & Secondary & Sun\\
\hline
\vspace*{-3mm}\\
He/H (nb)       & $0.0851^a$ & $0.0851^a$ & $0.0851\pm0.0020$ \\
\vspace*{-3mm}\\
C/H (nb) & $1.17^{+0.30}_{-0.27}\times10^{-4}$ & $2.12^{+0.31}_{-0.28}\times10^{-4}$ & $2.69\pm0.31\times10^{-4}$ \\ 
\vspace*{-3mm}\\
N/H (nb) & $1.32^{+0.90}_{-0.72}\times10^{-4}$ & $1.15^{+0.69}_{-0.55}\times10^{-4}$ & $6.76\pm0.78\times10^{-5}$\\ 
\vspace*{-3mm}\\
O/H (nb) & $4.29^{+1.07}_{-0.70}\times10^{-4}$ & $6.01^{+1.81}_{-1.21}\times10^{-4}$ & $4.90\pm0.56\times10^{-4}$\\ 
\vspace*{-3mm}\\
\hline
\end{tabular}
\end{center}
\tablefoot{ $^a$Value fixed to the solar value \citep{Asplund}. }
\label{Table:CMFGEN_ab}
\end{table*} 

The bolometric magnitudes of the stars $M_{\text{bol},1}=M_{\text{bol},2}=-8.63^{+0.26}_{-0.23}$ were computed assuming that the bolometric correction depends only on the effective temperature through the relation 
\begin{equation}
BC = -6.89\log(\text{T}_{\text{eff}}) + 28.07
\end{equation}
\citep{MartinsPlez}. We further obtained bolometric luminosities $L_\text{bol}$ of $2.27^{+0.55}_{-0.47}\times 10^5 L_\odot$ for both stars. From the relation between bolometric luminosity and radius, we infer a spectroscopic radius $R_\text{spectro}=13.7^{+1.9}_{-1.6} \,\text{R}_\odot$ for both stars. Using the surface gravity determined with {\tt CMFGEN}, we obtain a spectroscopic mass $\text{M}_\text{spectro}=20.8^{+7.4}_{-6.9}\,\text{M}_\odot$ for both stars. \\

It is important to recall that {\tt CMFGEN} assumes that the stars are spherical, static, and isolated. However, because of the binarity, these three assumptions are not verified. The effective temperature is thus not homogeneous all over the surface and the computed temperature has to be considered as a mean value over the star. In addition, because of the companion, the surface gravity $|\vec{\nabla}(\Omega)|$ is not homogeneous over the stellar surface and the local surface gravity is lower than that of an isolated star. As a consequence, the values of $\log g$ computed with {\tt CMFGEN} lead to spectroscopic masses that underestimate the actual masses.

\section{Radial velocity analysis \label{omegadot}}
Combining our new RVs determined through spectral disentangling with the literature data from \citet{Struve}, \citet{Hill}, \citet{Penny}, the B\&C and CAT/CES data from \citet{Sana01}, and the data from \citet{Mayer08}, we obtain 143 RV measurements spread over 67 years. For the RV data from \citet{Struve} and \citet{Hill}, we estimated errors on individual data points of 15\,km\,s$^{-1}$. For the Cerro-Tololo, Echelec, and CAT/CES data of \citet{Mayer08}, we estimated errors of 8\,km\,s$^{-1}$, 5\,km\,s$^{-1}$ and 5\,km\,s$^{-1}$, respectively. For the {\it IUE} data from \citet{Penny} and the B\&C RVs from \citet{Sana01}, we adopted an error estimate of 10\,km\,s$^{-1}$. Finally, for the CAT/CES data of \citet{Sana01} and the RVs derived from our spectral disentangling, we took 5\,km\,s$^{-1}$ and 3\,km\,s$^{-1}$, respectively.

For an SB2 binary such as HD~152\,248, the primary and secondary radial velocities are given by
\begin{equation}
  RV_{\rm P}(t) = K_{\rm P}\,(\cos{(\phi(t)+\omega(t))}+e\,\cos{\omega(t)})+\gamma_{\rm P},
\end{equation}
and
\begin{equation}
  RV_{\rm S}(t) = -K_{\rm S}\,(\cos{(\phi(t)+\omega(t))}+e\,\cos{\omega(t)})+\gamma_{\rm S},
\end{equation}
where $K_{\rm P}$ and $K_{\rm S}$ are the amplitudes of the RV curves, and $\phi(t)$ is the true anomaly computed with Kepler's equation with a specific value of the anomalistic orbital period (time interval between two consecutive periastron passages) $P_{\rm orb}$, the eccentricity $e,$ and the time of periastron passage $T_0$. Likewise, $\omega(t)$ is the longitude of periastron of the primary star at time $t,$ and the $\gamma_{\rm P}$ and $\gamma_{\rm S}$ are the apparent systemic velocities. Therefore we can estimate the mass-ratio of the binary by fitting a linear relation
\begin{equation}\label{massratio}
  RV_{\rm P}(t) = -q\,RV_{\rm S}(t)+B,
\end{equation}
where $q = \frac{K_{\rm P}}{K_{\rm S}} = \frac{m_{\rm S}}{m_{\rm P}}$ and $B = \gamma_{\rm P} + q\,\gamma_{\rm S}$.
Applying this linear regression to the various datasets, we obtained $q = \frac{m_{\rm S}}{m_{\rm P}} = 1.04 \pm 0.03$. Because each literature source of RV data has its own convention for the apparent systemic velocities, we found different values of the $B$ parameter for each dataset. 
Equation\,\ref{massratio} was then used to convert the RVs of both stars into equivalent RVs of the primary with
\begin{equation}\label{RVeq}
  RV_{\rm eq}(t) = (RV_{\rm P}(t)-q\,RV_{\rm S}(t)+B)/2.
\end{equation}
For each time of observation $t$, we adjusted the primary, secondary, and equivalent RV data with the following relation
\begin{equation}
\label{eq:RV}
RV(t) = \gamma + K\,[\cos{(\phi(t)+\omega(t))} + e\,\cos{\omega(t)}],
\end{equation}
where 
\begin{equation}
\label{eq:w}
\omega(t) = \omega_0 + \dot{\omega}\,(t - T_0).
\end{equation}
$\omega_0$ is the value of $\omega$ at time $T_0$ and $\dot{\omega}$ is the rate of secular apsidal motion. The apparent systemic velocity was adjusted for each RV dataset so as to minimise the sum of the residuals of the data about the curve given by Eq.\,\ref{eq:RV}. We tested different values of $K_{\rm P}$ ranging from 213 to 219\,km\,s$^{-1}$. The best quality of the fit, as judged from its global $\chi^2$, was obtained for $216^{+2}_{-1}$\,km\,s$^{-1}$. We accordingly adopted this value for the amplitude of variations of the primary and equivalent RVs. Likewise, we tested different values of $K_{\rm S}$ between 209 and 216\,km\,s$^{-1}$, and the best results were obtained for $K_{\rm S} = 212^{+2}_{-2}$\,km\,s$^{-1}$. For the equivalent RVs, we finally determined the value of $q = 1.01^{+0.02}_{-0.01}$ , which yields the best global fit to the data through Eq.\,\ref{eq:RV}. We therefore conclude that the mass ratio is consistent with 1.0. 

\begin{figure}[h]
  \begin{center}
    \begin{minipage}{4cm}
      \includegraphics[width=4cm]{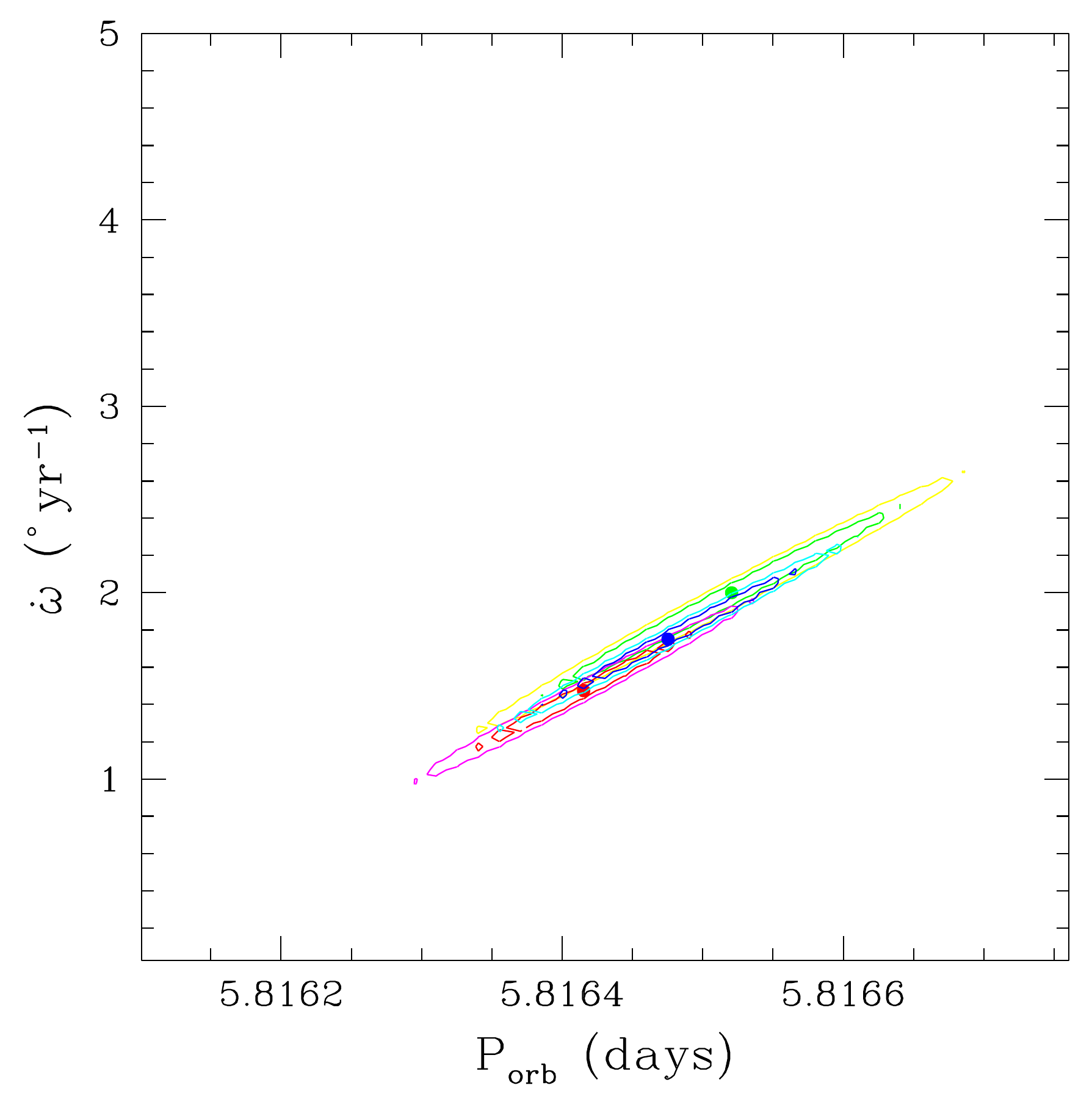}
    \end{minipage}
    \begin{minipage}{4cm}
      \includegraphics[width=4cm]{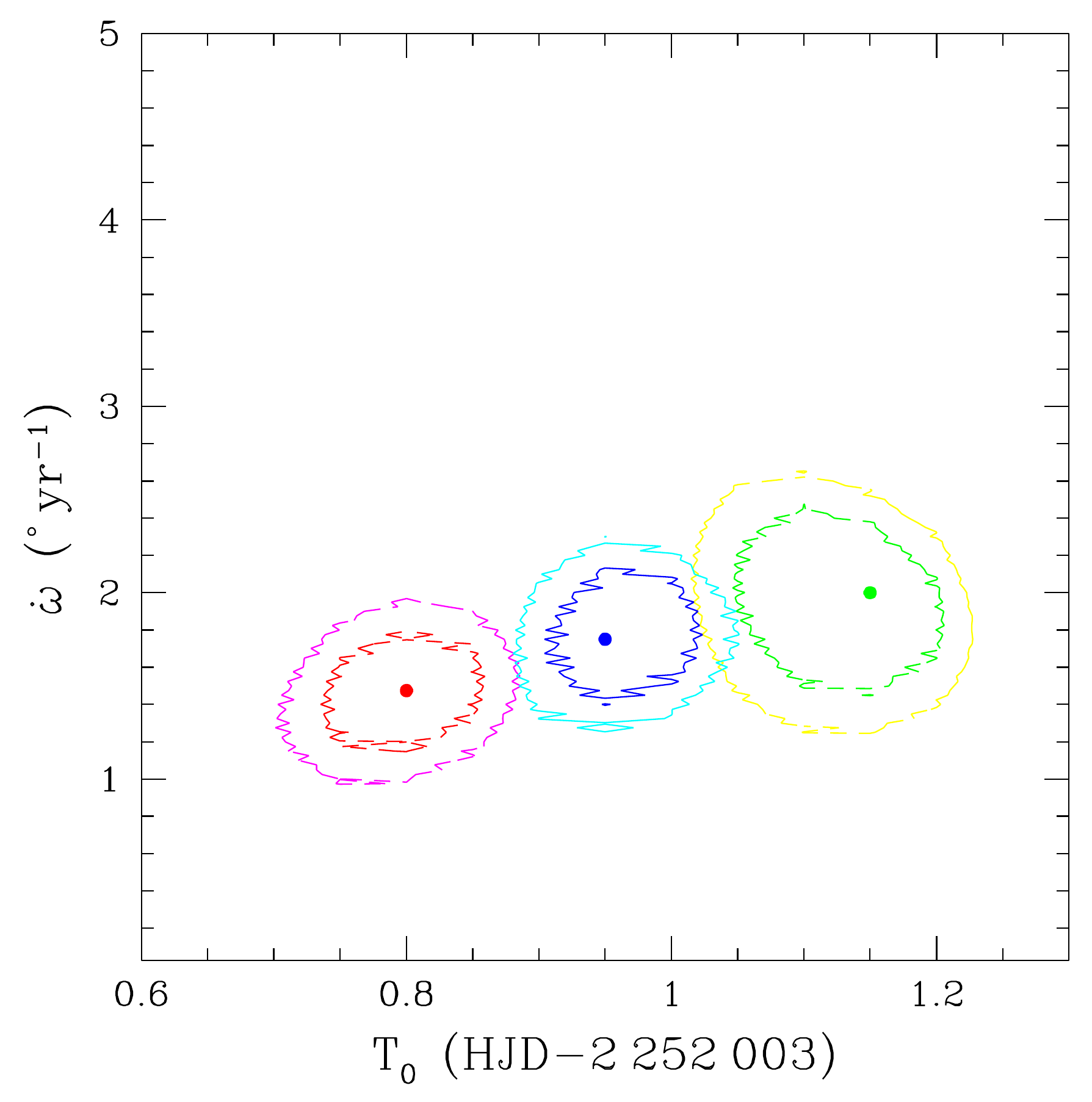}
    \end{minipage}
    \begin{minipage}{4cm}
      \includegraphics[width=4cm]{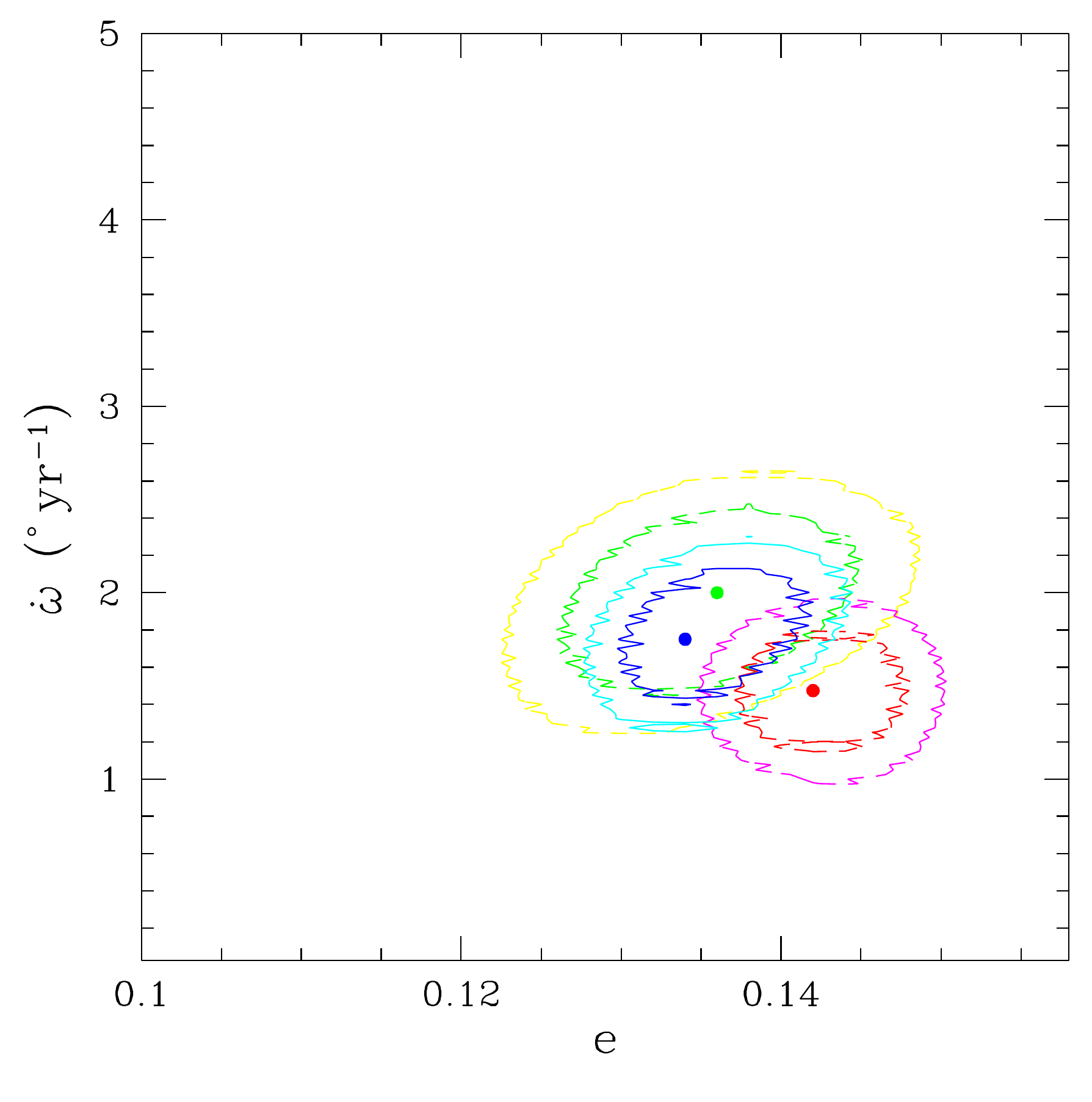}
    \end{minipage}
    \begin{minipage}{4cm}
      \includegraphics[width=4cm]{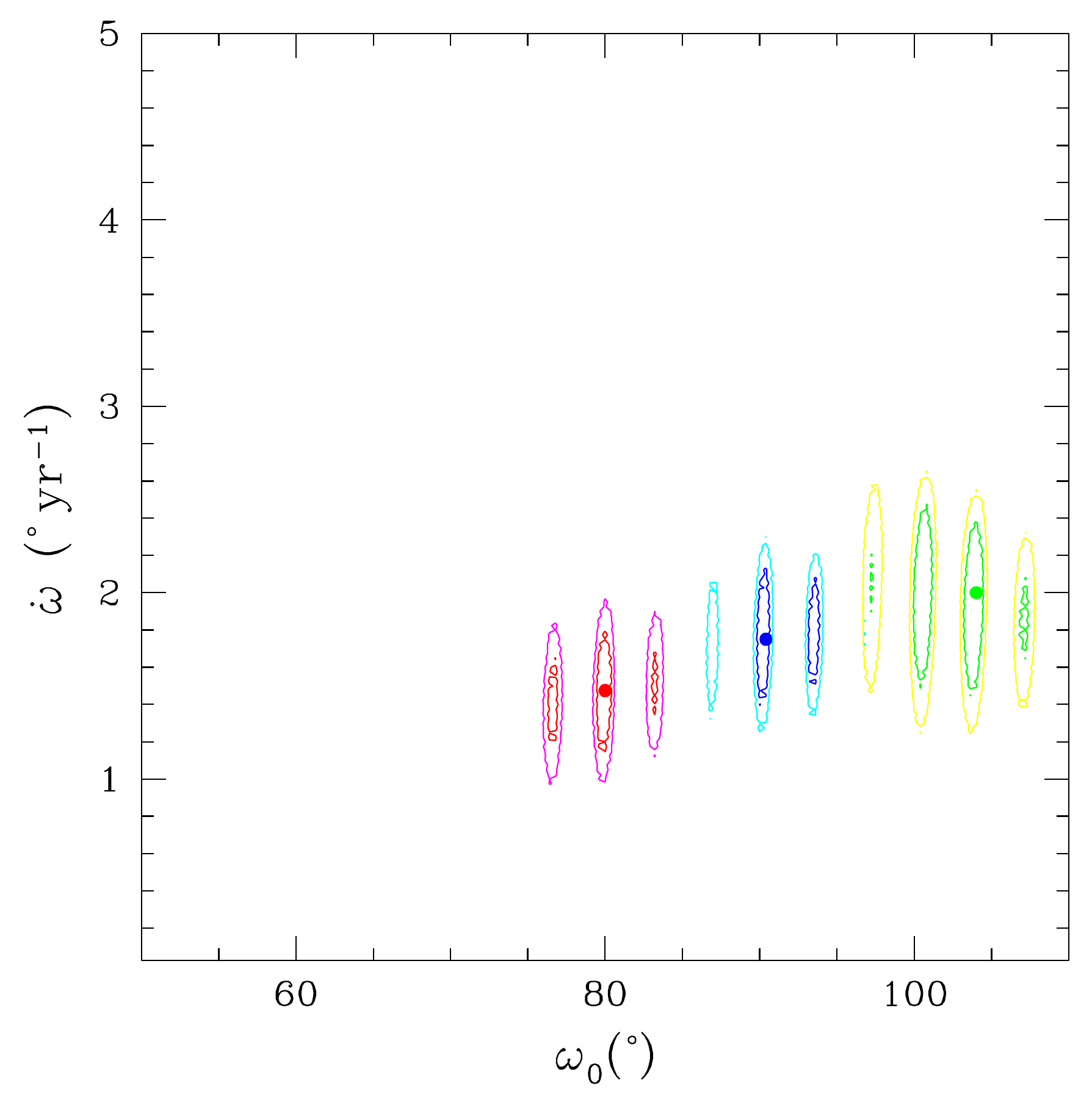}
    \end{minipage}
  \end{center}
  \caption{Confidence contours for the best-fit parameters obtained from the adjustment of the full set of 143 RV data with Eqs.\,\ref{eq:RV} and \ref{eq:w}. The best-fit solutions for the primary, secondary, and equivalent RVs are shown by the red, green, and blue filled dots. The corresponding $1\,\sigma$ and 90\% confidence levels are shown by the red and magenta, green and yellow, and blue and cyan contours for the primary, secondary, and equivalent RVs, respectively.\label{contoursomega}}
\end{figure}

The problem then consists of finding the five free parameters ($P_{\rm orb}$, $e$, $T_0$, $\omega_0$ , and $\dot{\omega}$) that provide the best fit to the whole set of RV data. We scanned the parameter space in a systematic way. The most important projections of the 5D parameter space onto 2D planes are shown in Fig.\,\ref{contoursomega}. We obtained somewhat different values of $\omega_0$ and $T_0$ depending on the RV sets (primary, secondary, or equivalent) that were considered. However, the rate of apsidal motion $\dot{\omega}$, the period $P_{\rm orb}$ , and to a lesser extent, the eccentricity $e$ are consistent for all datasets.

\begin{figure}[htb]
\begin{center}
\includegraphics[width=8cm]{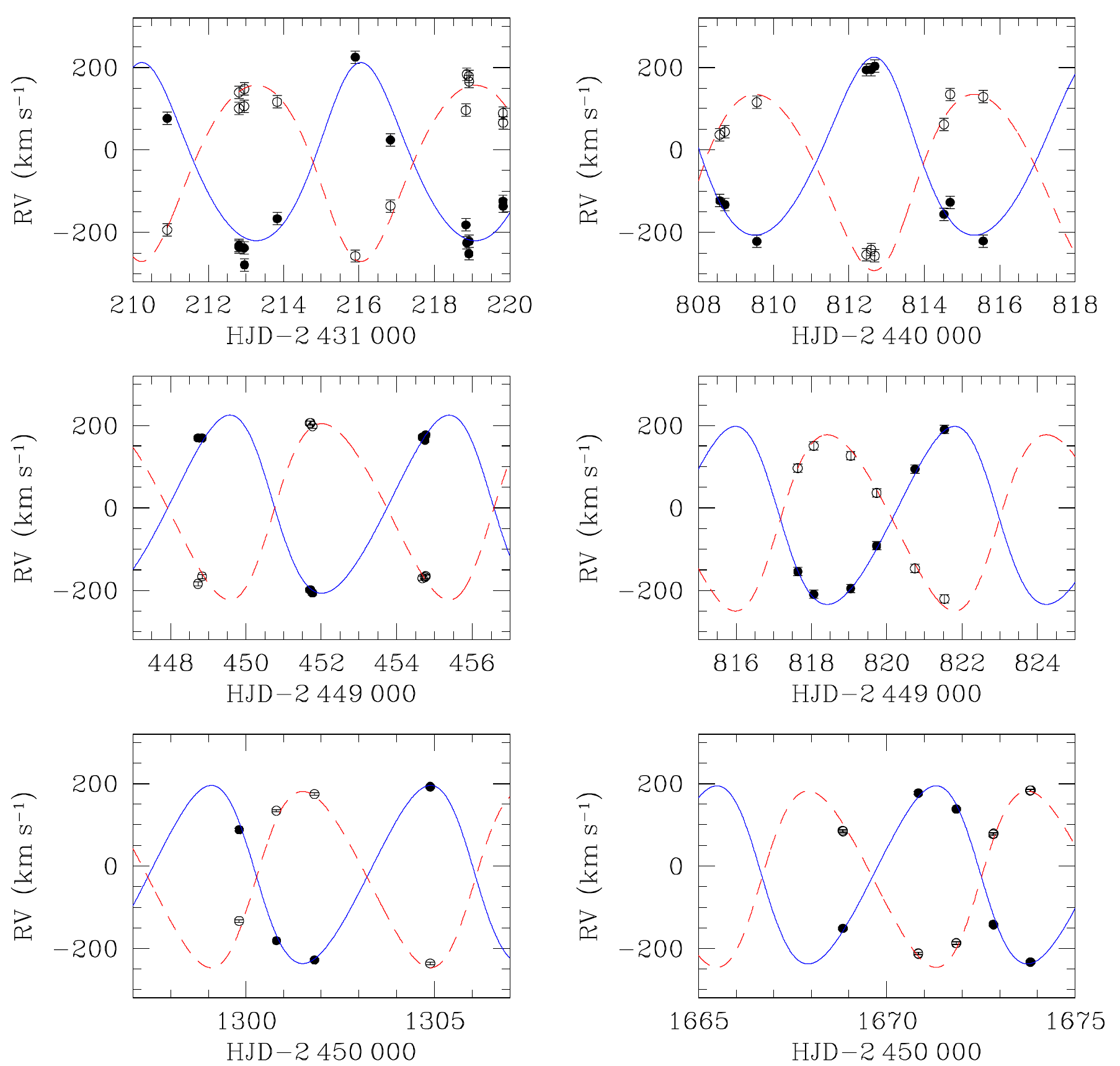}
\end{center}
\caption{Comparison between the measured RVs of the primary (filled dots) and secondary (open dots) and the RVs expected from Eqs.\,\ref{eq:RV} and \ref{eq:w} with the best-fit parameters given in Table\,\ref{bestfitTable}. The top panels correspond to data from \citet[][left]{Struve} and \citet[][right]{Hill}. The middle panels show one epoch from \citet[][left]{Mayer08}, as well as one epoch of {\it IUE} data from \citet[][right]{Penny}. The bottom panels correspond to RVs that we rederived here. \label{fitRV}}
\end{figure}

In this way, we found the best-fit parameters and their $1\sigma$ errors, which are given in Table\,\ref{bestfitTable}. The best fit of the RV data at six different epochs is illustrated in Fig.\,\ref{fitRV}. The change in morphology of the RV curve over more than six decades between the \citet{Struve} data and the most recent observations is clearly visible. We thus find that the value of $\dot{\omega}$ that best fits our RV time series is $(1.750^{+0.350}_{-0.315})^{\circ}$\,yr$^{-1}$. We stress that the uncertainties account for correlations that exist between the different parameters appearing in the fitting relation (see Fig.\,\ref{contoursomega}).

\begin{table*}[htb]
\caption{Best-fit orbital parameters of HD~152\,248 obtained with Eqs.\,\ref{eq:RV} and \ref{eq:w}.}
\begin{center}
\begin{tabular}{l l l l}
\hline\hline
Parameter & Primary RVs & Secondary RVs & Equivalent RVs\\
\hline
\vspace*{-3mm}\\
$P_{\rm orb}$\,(d) & $5.816415^{+0.000085}_{-0.000075}$ & $5.816520^{+0.000130}_{-0.000120}$ & $5.816475^{+0.000085}_{-0.000075}$ \\
\vspace*{-3mm}\\
$e$              & $0.142^{+0.005}_{-0.004}$ & $0.136^{+0.009}_{-0.010}$ & $0.134^{+0.007}_{-0.004}$\\
\vspace*{-3mm}\\
$\dot{\omega}$\,($^{\circ}$\,yr$^{-1}$) & {\bf $1.48^{+0.36}_{-0.31}$} & {\bf $2.00^{+0.48}_{-0.53}$} & {\bf $1.75^{+0.35}_{-0.32}$} \\
\vspace*{-3mm}\\
$\omega_0$\,($^{\circ}$) & $80.0^{+3.6}_{-3.6}$ & $104.0^{+3.6}_{-6.7}$ & $90.4^{+3.6}_{-3.3}$ \\
\vspace*{-3mm}\\
$T_0$\,(HJD$-$2\,450\,000) & $2\,003.80^{+0.05}_{-0.06}$ & $2\,004.15^{+0.05}_{-0.10}$ & $2\,003.95^{+0.07}_{-0.04}$\\
\vspace*{-3mm}\\
$q = m_2/m_1$   & ... & ...  &  $1.01^{+0.02}_{-0.01}$  \\
\vspace*{-3mm}\\
$K_1$\,(km\,s$^{-1}$) & $216^{+2}_{-1}$ & ... & $216$ (adopted)\\
\vspace*{-3mm}\\
$K_2$\,(km\,s$^{-1}$) & ... & $212^{+2}_{-2}$ & ... \\
\vspace*{-3mm}\\
$a_1\,\sin{i}\,(\text{R}_{\odot})$ & $24.6^{+0.2}_{-0.1}$ & ... & $24.6^{+0.2}_{-0.1}$ \\
\vspace*{-3mm}\\
$a_2\,\sin{i}\,(\text{R}_{\odot})$ & ... & $24.1^{+0.2}_{-0.2}$ & $24.4^{+0.2}_{-0.2}$\\
\vspace*{-3mm}\\
$m_1\,\sin^3{i}\,(\text{M}_{\odot})$ & ... & ... & $23.2^{+0.7}_{-0.6}$ \\
\vspace*{-3mm}\\
$m_2\,\sin^3{i}\,(\text{M}_{\odot})$ & ... & ... & $23.4^{+0.7}_{-0.4}$\\
\vspace*{-3mm}\\
$\chi^2_\nu$ & 2.831 & 3.967 & 2.082 \\     
\vspace*{-3mm}\\
\hline
\end{tabular}
\end{center}
\label{bestfitTable}
\end{table*} 

\citet{Mayer08} derived a rate of apsidal motion of $(2.13 \pm 0.23)^{\circ}$\,yr$^{-1}$ from the full set of RVs available to them. Within the error bars, their and our values overlap. However, \citet{Mayer08} favoured a higher value of $(3.07 \pm 0.47)^{\circ}$\,yr$^{-1}$, obtained by discarding the \citet{Struve} measurements. We therefore repeated some calculations of $\dot{\omega}$ discarding at first the \citet{Struve}, and in a second step also the \citet{Hill} data. Our best-fit results lead to a reduction by nearly 30\% of $\dot{\omega}$ compared to our favoured value, instead of an increase as advocated by \citet{Mayer08}. At the same time, the uncertainties increase because of the significantly reduced time spanned by the dataset. We therefore decided to keep all the RV data in our analysis.   

To verify the consistency of our determination of $\dot{\omega}$, we determined the values of $\omega$ at all epochs for which at least 5 RV data points were available. In this process, the eccentricity, the anomalistic orbital period, the semi-amplitudes of the primary RV curve, and the mass ratio were frozen to the values obtained from the analysis of the equivalent RVs ($e = 0.134$, $P_{\rm orb} = 5.816475$\,d, $K_1 = 216$\,km\,s$^{-1}$, $q = 1.01$). For each subset of RV data, the systemic velocity and the $\omega$ value were then adjusted in order to achieve the smallest $\chi^2$ between the observed and computed RVs. The results agree very well with the $\omega$ values obtained from the global fit of the equivalent RVs. While some scatter exists around the linear $\omega(t) = \omega_0 + \dot{\omega}\,(t-T_0)$ relation, we find no compelling evidence for a more complex dependence of $\omega$ on time. The scatter is fully consistent with the errors on the individual $\omega$ determinations and with the uncertainties of the global fit to the equivalent RVs. Further aspects of the RVs of HD~152\,248, notably regarding the possibility of a triple system, are discussed in Appendix~\ref{appendix}.

\section{Light curve analysis\label{photom}}
The $c_2$ light curve of HD~152\,248 displays minima of unequal depths. In the present case, this difference in depth does not arise from a difference in surface brightness of the two stars, but rather from the fact that the orbit is eccentric and that $\omega$ was close to $90^{\circ}$ at the time of our photometric campaign. 
We analysed the light curves of HD~152\,248 with the {\tt Nightfall} code (version 1.86) developed by R.\ Wichmann, M.\ Kuster, and P.\ Risse\footnote{The code is available at the URL: http://www.hs.uni-hamburg.de/DE/Ins/Per/Wichmann/Nightfall.html} \citep{Wichmann}. This code uses the Roche potential scaled with the instantaneous separation between the stars to describe the shape of the stars. For an eccentric binary without surface spots, the model is described by eight parameters: the mass ratio $q$, the orbital inclination $i$, the primary and secondary filling factors ($f_{\rm P}$ and $f_{\rm S}$, defined as the ratio between the stellar polar radius and the polar radius of the associated instantaneous Roche lobe at periastron), the primary and secondary effective temperatures (T$_{\rm eff,P}$ and T$_{\rm eff,S}$), the eccentricity $e$ and the argument of periastron $\omega$. Following the results of our analysis of the RVs (see Sect.\,\ref{omegadot}), we set $q$ to 1.0 and $e$ to $0.134$. The stellar effective temperatures of both stars were fixed to 34\,000\,K as derived from our {\tt CMFGEN} analysis (Sect.\,\ref{fitcmfgen}). We adopted a quadratic limb-darkening law. Reflection effects were accounted for by considering the mutual irradiation of all pairs of surface elements of the two stars \citep{Hendry}. Leaving $f_{\rm P}$ and $f_{\rm S}$ unconstrained yields solutions that are often at odds with the spectroscopic brightness ratio of 1.0. Therefore we chose to adopt the condition that $f_{\rm P} = f_{\rm S}$. We thus searched for the values of $\omega$, $i$ and $f_{\rm P} = f_{\rm S}$ that yield the best fit to the observed light curves. We started with the $c_2$-continuum light curve because the He\,{\sc ii} $\lambda$\,4686 line-bearing filter could be affected by phase-dependent emission from the interaction zone of the stellar winds \citep{Sana01}.

\begin{figure}[htb]
  \begin{center}
  \resizebox{8cm}{!}{\includegraphics{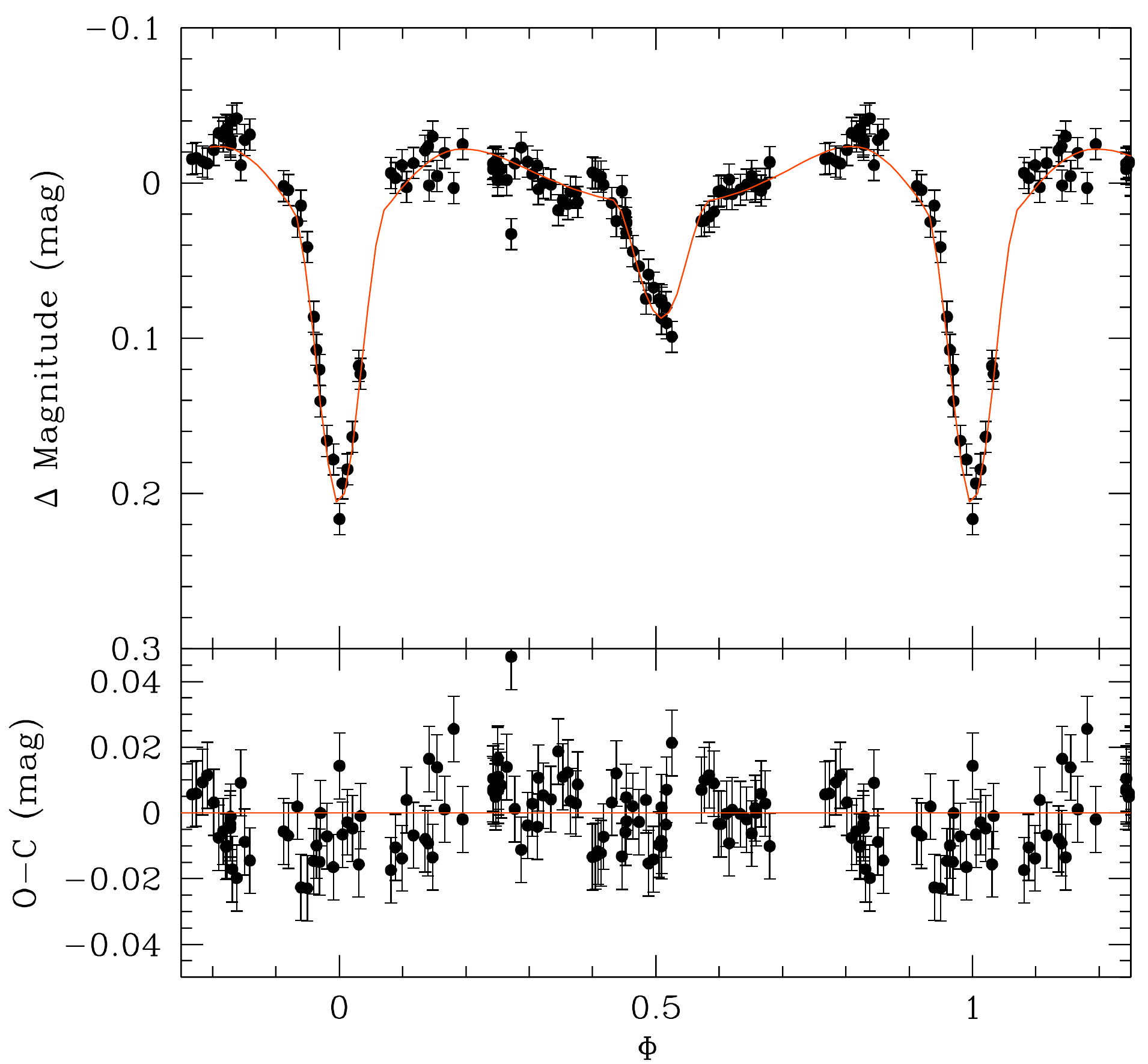}}
\end{center}  
\caption{Best-fit {\tt Nightfall} solution of the $c_2$-band light curve of HD~152\,248. The lower panel shows the residuals over the best-fit solution.\label{bestfit}}
\end{figure}

With 125 data points that leave very few gaps in orbital phase, our dataset provides the best coverage of the orbital cycle at a specific epoch among all existing light curves of HD~152\,248. Table\,\ref{parambestfit} lists the best-fit parameters of the $c_2$-band light curve. The best fit is illustrated in Fig.\,\ref{bestfit}. The dispersion of the data points about the theoretical light curve exceeds the formal photometric errors. The same observation can be made on the photometric data of \citet{Mayer08}, indicating that the additional variability is most probably intrinsic to the stars that make up the binary system and is not of instrumental origin. This is a relatively common situation for light curves of massive stars, which frequently display low-level intrinsic photometric variations \citep[see, e.g.,\ the red noise variability of the O-star binary HD~149\,404,][]{Rauw19}. It is worth noting that the light curve taken with the filter centred on the He\,{\sc ii} $\lambda$\,4686 line is in excellent agreement with the best-fit $c_2$-band curve, and does not reveal any strange behaviour due to the phase-dependent He\,{\sc ii} $\lambda$\,4686 emission.    

\begin{table}
  \caption{Parameters of the best-fit {\tt Nightfall} model for the $c_2$ light curve. \label{parambestfit}}
  \begin{center}
  \begin{tabular}{l l}
    \hline\hline
    Parameter & Value \\
    \hline
    T$_{\rm eff,P}$ = T$_{\rm eff, S}$\,(K) & 34\,000 (fixed) \\
    $e$ & 0.134 (fixed) \\
    \vspace*{-3mm}\\
    $f_{\rm P} = f_{\rm S}$ & $0.86^{+0.03}_{-0.02}$ \\
    \vspace*{-3mm}\\
    $i$\,($^{\circ}$) & $68.6^{+0.2}_{-0.3}$ \\
    \vspace*{-3mm}\\
    $\omega$\,($^{\circ}$) & $84.7^{+2.4}_{-2.7}$ \\
    \vspace*{-3mm}\\
    $T_{\rm prim. ecl.}$ (HJD - 2\,450\,000) & $549.92$ \\
    \vspace*{-3mm}\\
    $\chi^2_\nu$ & 1.395 \\     
    \hline
  \end{tabular}
  \end{center}
\end{table}

The light curves of eclipsing binaries are also quite sensitive to apsidal motion and are frequently used to establish the presence of this phenomenon \citep[e.g.,][]{Zasche}. In the case of HD~152\,248, the eclipses were discovered and confirmed well after the discovery of the SB2 nature of the system \citep{Mayer92}. The total time interval over which photometric data are available is therefore relatively short ($\sim 27$\,yr) compared to the duration over which spectroscopic data were collected (67\,yr). Therefore, we expect the determination of $\dot{\omega}$ based on photometric data to be more difficult than via spectroscopy.

Nevertheless, the existing photometric data allow us at least to perform a consistency check of the results derived from the spectroscopic data. To do so, we followed two different avenues.

As a first step, we analysed the three epochs of $UBV$ photometry from \citet{Mayer08} that allowed us to build a reasonably well-sampled light curve collected over a duration shorter than a month. These subsamples of the \citet{Mayer08} data were obtained in March 1992, June 1993, and April-May 2004. The remainder of their data form subsamples that are too scarce to derive meaningful constraints on $\omega$ for a given individual epoch. The light curves from the three epochs were fit with {\tt Nightfall} setting the Roche-lobe filling factors, the inclination, and the eccentricity to the best-fit values of the $c_2$ light curve (see Table\,\ref{parambestfit}), and keeping $\omega$ as the only free parameter. The values of $\omega$ obtained this way are listed in Table\,\ref{omegaMayer}.

\begin{table}
  \caption{Best-fit values of $\omega$ from photometry.}
  \begin{center}
  \begin{tabular}{c c c c c}
    \hline\hline
    Epoch & $N$ & $\Delta t$ & Filter & $\omega$ \\
    HJD-2\,440\,000 & & (d) & & $(^{\circ})$ \\
    \hline
    8686.5 & 234 & 9.05 & $UBV$ & $70.4 \pm 1.2$ \\
    9157.0 & 246 & 9.92 & $UBV$ & $83.6 \pm 1.2$ \\
    13123.5 & 220 & 11.20 & $UBV$ & $102.3 \pm 2.1$ \\
    \vspace{-2mm}\\
    10545.0 & 125 & 27.97 & $c_2$ & $84.7 \pm 2.6$ \\
     \vspace{-2mm}\\
  18641.4 & 12\,372 & 23.04 & {\it TESS} & $128.7 \pm 0.9$ \\
    \hline
  \end{tabular}
  \end{center}
  \tablefoot{The first three lines correspond to the fits of the $UBV$ photometry from \citet{Mayer08}, and the fourth and last line correspond to the fit of the $c_2$ light curve and the \textit{TESS} photometry, respectively. $N$ is the number of photometric data points, and $\Delta t$ expresses the duration of the observing campaign.\label{omegaMayer}}
\end{table}

Regarding the {\it TESS} data, the features mentioned in Sect.\,\ref{sect:photo} prevent us from deriving an independent photometric solution from this dataset. Instead, we adopted the same approach as for the \citet{Mayer08} photometry, fixing all parameters except for $\omega$ to their values inferred from our $c_2$ light curve. We then searched for the best-fit value of $\omega$ using either the two time intervals of {\it TESS} data separately or combining them into a single dataset. In this way, we derived $\omega = 128.7 \pm 0.9$ as the best estimate of the longitude of periastron at the epoch of the {\it TESS} observations. 

\begin{figure}[htb]
  \begin{center}
  \resizebox{9cm}{!}{\includegraphics{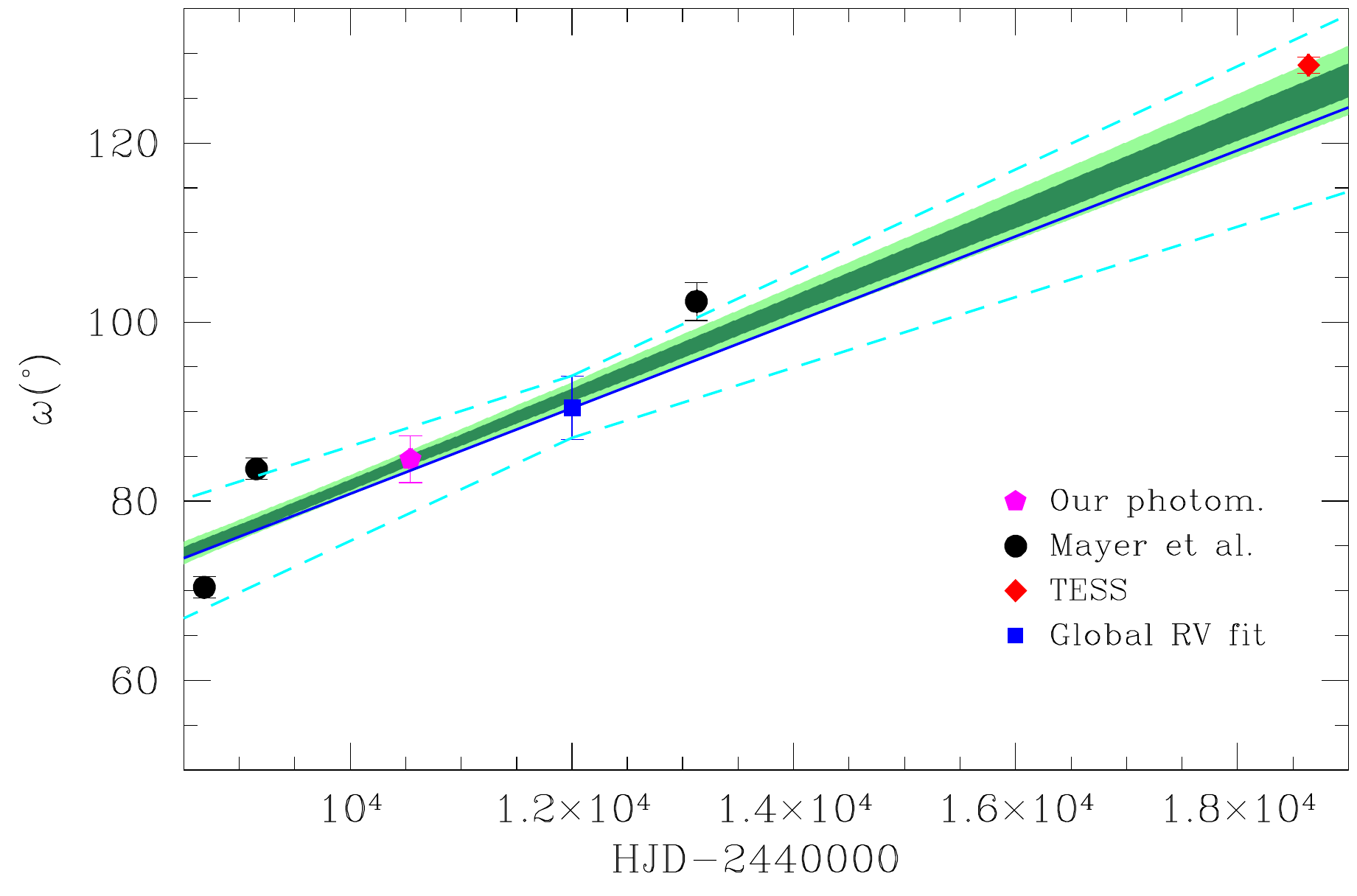}}
\end{center}  
\caption{Values of $\omega$ as a function of time inferred from the photometric light curves and the RVs. The black filled dots correspond to the data of the fits of the $UBV$ photometry from \citet{Mayer08} (the first three lines of Table\,\ref{omegaMayer}), the pink filled dot corresponds to the fit of the $c_2$ light curve (penultimate line of Table\,\ref{omegaMayer}), the red filled square corresponds to the data of the fits of the {\it TESS} photometry (last line of Table\,\ref{omegaMayer}), and the blue filled square indicates the $\omega_0$ value obtained from the global fit of all RV data. The solid blue line corresponds to our best value of $\dot{\omega}$ from the equivalent RVs, and the dashed cyan lines corresponds to the range of values according to the $1\,\sigma$ uncertainties on $\omega_0$ and $\dot{\omega}$ (Table\,\ref{bestfitTable}). In dark green and light green we represent the 1$\sigma$ and 2$\sigma$ uncertainties on $\omega(t)$ from the posteriors for $\omega$ and $\dot{\omega}$ as shown in Fig.\,\ref{fig:phoebe_corner} (see {\tt PHOEBE} analysis in Sect.~\ref{sect:phoebe}).   The {\tt PHOEBE} model is fit directly to the observations, not to the estimates shown in this figure.  \label{omegaphot}}
\end{figure}

Figure\,\ref{omegaphot} illustrates the variations of $\omega$ with time. While the overall trend of the photometric data agrees with the result inferred from the global fit of the RV data, we note that there is quite some dispersion of the photometrically inferred values compared to the relation derived from the RV data. While this dispersion could partially result from the intrinsic photometric variability that we have pointed out in Sect.\,\ref{photom}, the position of the June 1993   (epoch 2\,449\,157.0~HJD) seems difficult to reconcile with a linear relation between $\omega$ and the time, even when we account for intrinsic photometric variations. However, the $\omega$ value inferred from the {\it TESS} photometry is consistent with the rate of apsidal motion determined from the RV data, which lends further support to our result.

As a second step, we attempted to use a maximum of the available photometric data, spanning as long a total duration as possible. To ensure that the light curves of the different subdatasets have morphologies that are as similar as possible, except for the effects of the apsidal motion, we selected the $V$-band and Str\"omgren $y$ data from \citet{Mayer08}. These filters are closest in wavelength to the $c_2$ filter we used in our study. Combining the $V$, $y$ and $c_2$-band data, we obtain a dataset consisting of 517 measurements that are (very) unequally spread over 14.26\,years. We excluded the {\it TESS} data from this exercise because of the artifacts mentioned in Sect.\,\ref{sect:photo} and because of their large number, which would clearly cause them to dominate all other datasets.

Setting the Roche-lobe filling factors and the orbital inclination to the values found from the $c_2$ light curve (see Table\,\ref{parambestfit}) and the eccentricity to the value found from the equivalent RVs (see Table\,\ref{bestfitTable}), we computed a grid of synthetic light curves for different values of $\omega$, covering the full range of possible values given the domain of $\omega_0$ and $\dot{\omega}$ to explore. Using these synthetic light curves, we computed the $\chi^2$ of the comparison between observed and expected magnitudes for all times of observation and for a set of four parameters ($P_{\rm orb}$, $T_0$, $\omega_0$ and $\dot{\omega}$). Because of the apsidal motion, the instantaneous period $P_{\rm ecl}$, that is,\ the time interval between two primary eclipses\footnote{For values of $\omega$ around $90^{\circ}$, the primary eclipse of HD~152\,248 corresponds to the primary star being eclipsed by the secondary.}, is not equal to the anomalistic period $P_{\rm orb}$, but amounts to 
\begin{equation}
  \label{Pecl}
  P_{\rm ecl} = \left[1 - \frac{(1 - e^2)^{3/2}}{(1 + e\,\sin{\omega})^2}\,\frac{\dot{\omega}\,P_{\rm orb}}{2\,\pi}\right]\,P_{\rm orb}
,\end{equation}
as pointed out by \citet{Schmitt}. This correction was applied to the calculation of the orbital phases from the times of the individual photometric data points.

\begin{figure*}[htb]
  \begin{center}
    \resizebox{6cm}{!}{\includegraphics{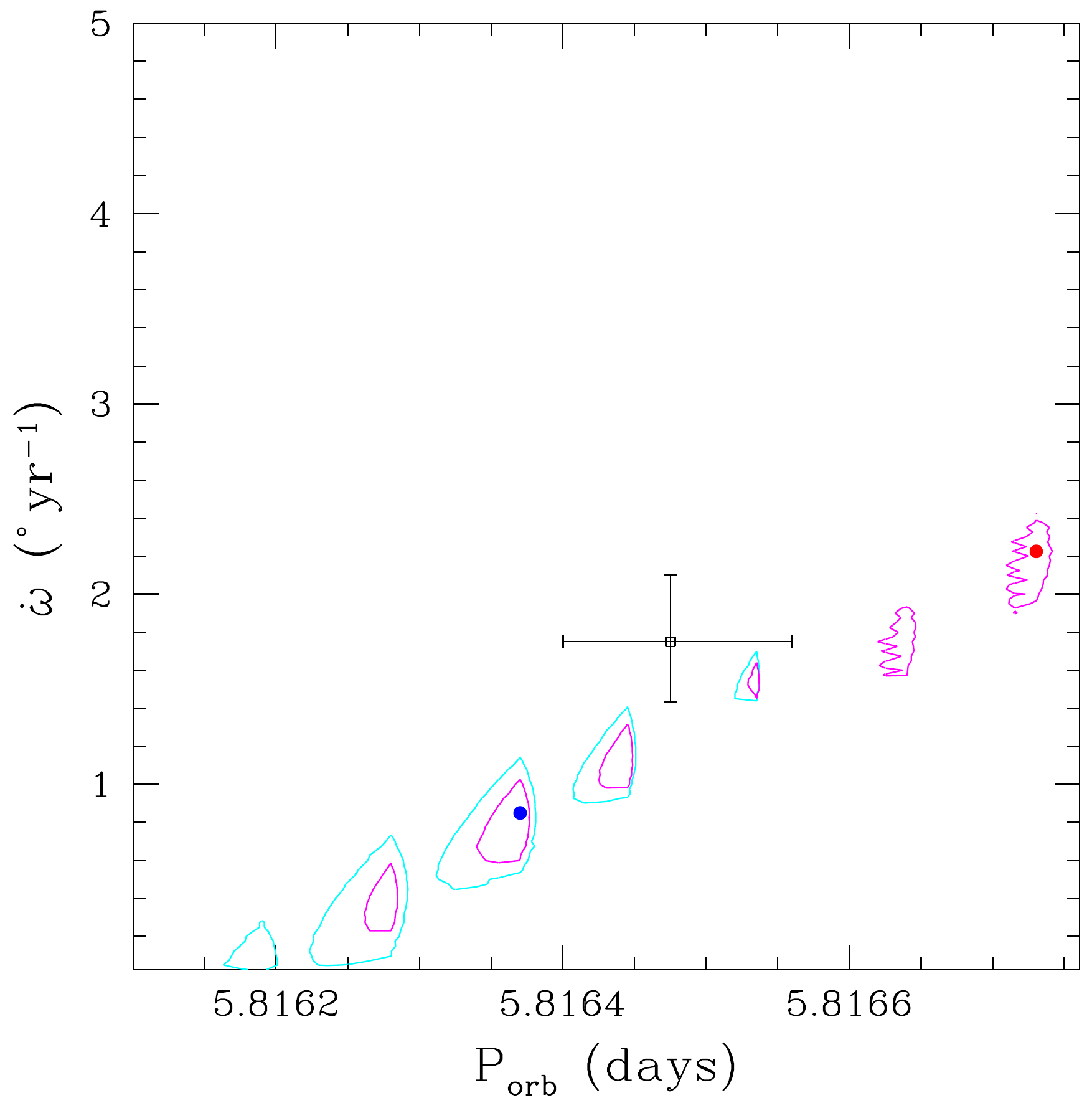}}
    \resizebox{6cm}{!}{\includegraphics{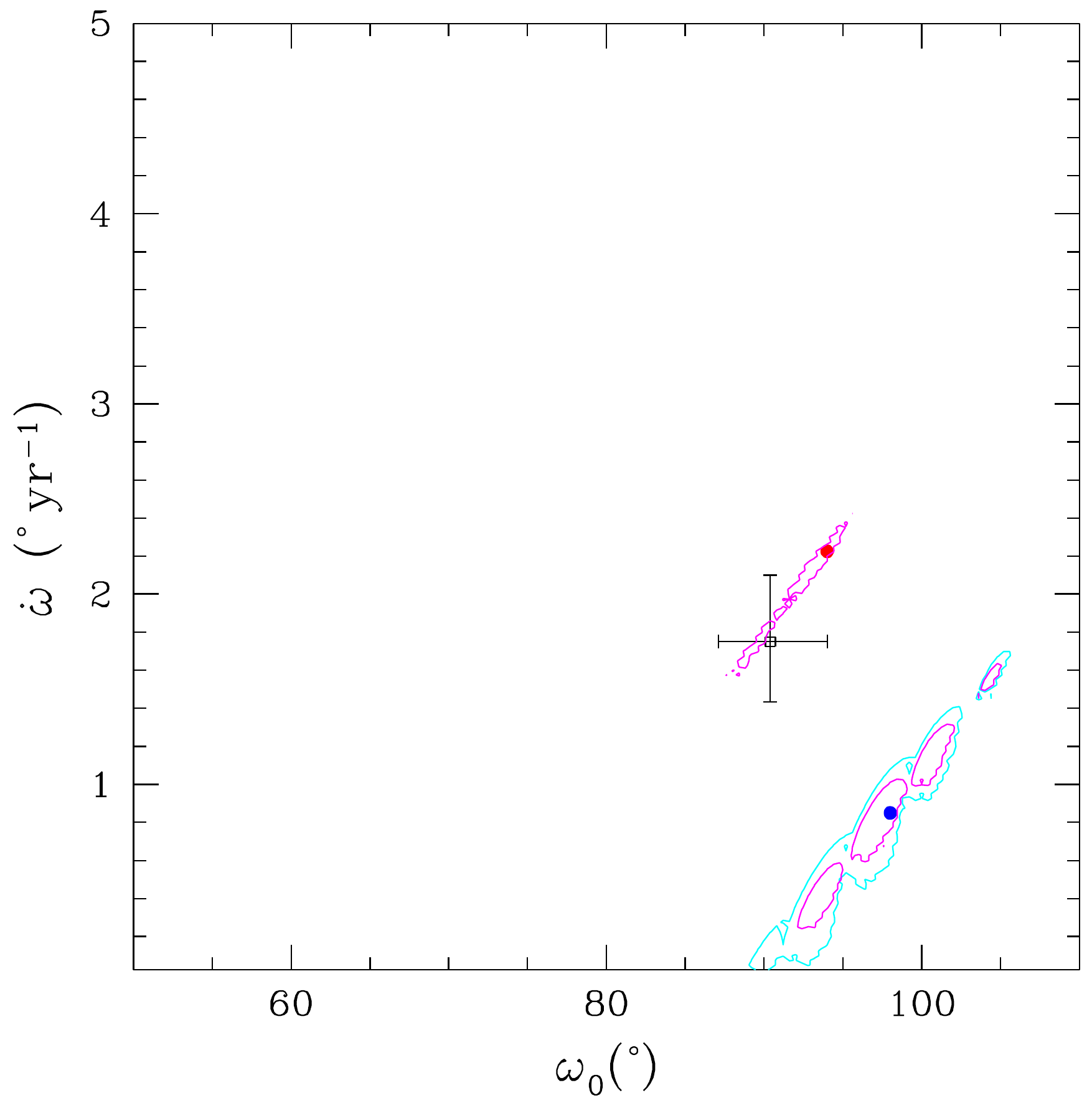}}
    \resizebox{6cm}{!}{\includegraphics{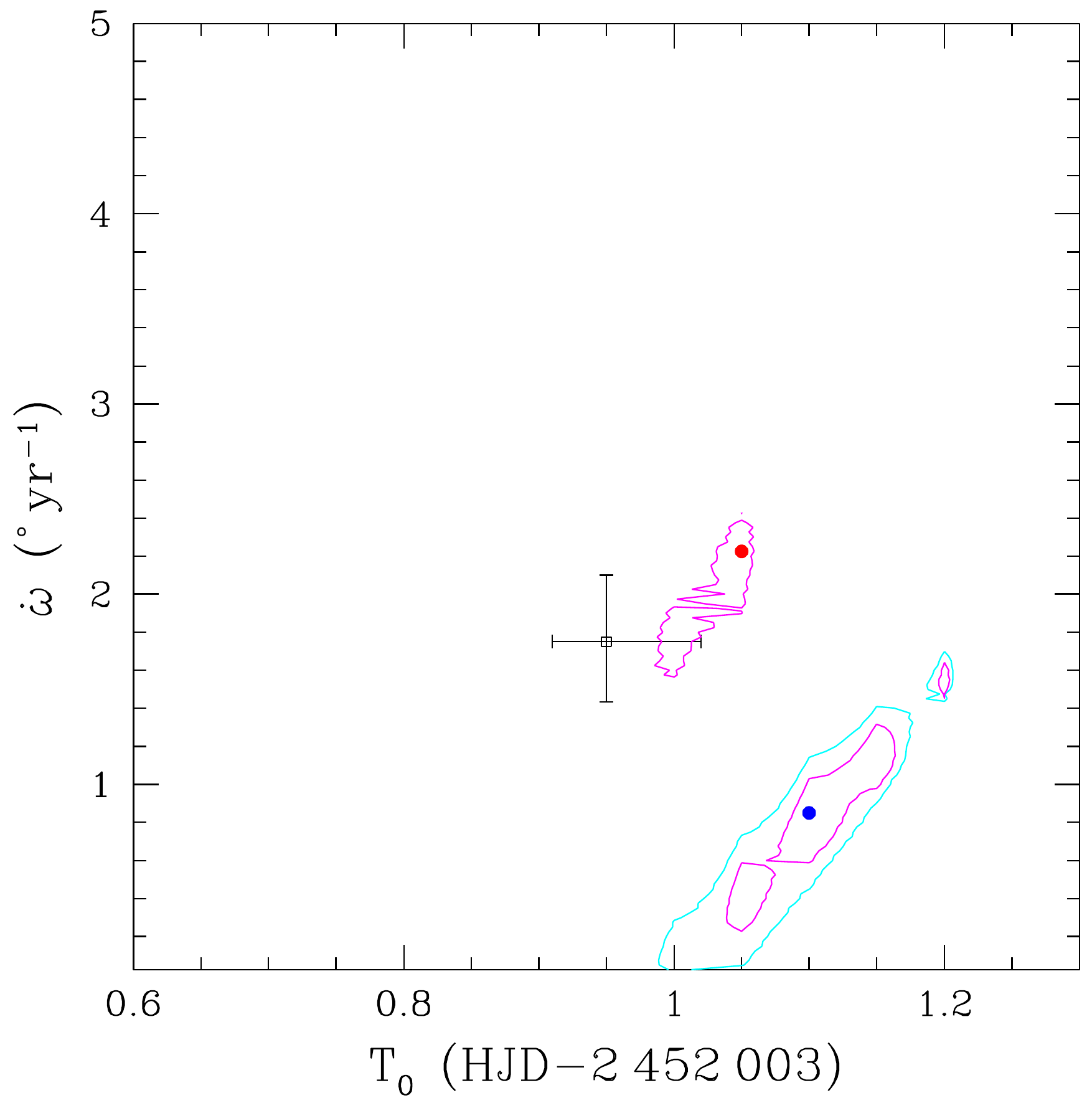}}
  \end{center}  
\caption{Contours of $\chi^2$ of the overall fits of the photometric data. The blue and red dots correspond  to the best fits for the entire set of $V$, $y$ and $c_2$-band data, and to the $V$ and $y$ data alone, respectively. The cyan and magenta contours provide the corresponding confidence levels  for the full dataset and for the sole $V$ and $y$ data, respectively. The open square with error bars corresponds to our best-fit result of the equivalent RV data.\label{photomcontours}}
\end{figure*}

The results are illustrated in Fig.\,\ref{photomcontours}.
As could be expected from the dispersion of photometrically inferred $\omega$ values in Fig.\,\ref{omegaphot}, the picture is far less clear than what we found from the RVs. For instance, applying the above method to the full 517 data points or restricting it to the 392 data points from \citet{Mayer08} yields very different results. In the first case, the best-fit $\dot{\omega}$ amounts to only $0.85^{\circ}$\,yr$^{-1}$, whereas in the second case, it reaches $2.23^{\circ}$\,yr$^{-1}$. Although the analysis of the individual photometric dataset reveals that $\omega$ changes from epoch to epoch, the currently available photometric data of HD~152\,248 do not allow us to independently establish a reliable rate of apsidal motion. 

\section{Complete {\bf \tt PHOEBE} binary model\label{sect:phoebe}}
The RV analysis (see Sect.\,\ref{omegadot}) and the light curve analysis (see Sect.\,\ref{photom}) were made independently, with each dataset being analysed with its own value of $\omega$ (see Figure 6). This  effectively allowed for apsidal motion between different datasets, but did not enforce any consistent linear rate of change in $\omega$.  Here we construct a comprehensive model from these independent analyses in {\tt PHOEBE~2} (version 2.2.0) \citep{Prsa, Jones}\footnote{{\tt PHOEBE~2} is an open-source project available at the URL http://phoebe-project.org.}. We determined robust posteriors using the affine-invariant Markov Chain Monte Carlo (MCMC) package, {\tt emcee} \citep{foreman-mackey}, on the photometry and RV data simultaneously with a single model and value for the rate of apsidal motion.  

{\tt PHOEBE~2} is an eclipsing binary modelling code that reproduces and fits light curves, RV curves, and spectral line profiles of eclipsing systems.  It includes advanced physics such as mutual irradiation, limb-darkening, light travel time effects, Doppler boosting, spin-orbit misalignment, and interstellar extinction.  Particularly relevant to this system, {\tt PHOEBE~2} handles apsidal motion by allowing the argument of periastron to vary by a linear rate of change in time.  In contrast to {\tt Nightfall}, {\tt PHOEBE~2} parameterises each component according to its equivalent (volumetric) radius  instead of filling factor or equipotential.  For eccentric orbits or misaligned systems, {\tt PHOEBE~2} then makes the assumption that the surface of the star instantaneously reacts to a change in the gravitational potential in a way that conserves the volume of each star. 

{\tt PHOEBE~2} works in flux units, therefore we converted the observed differential magnitudes into relative fluxes and allowed the synthetic light curve model to automatically scale to these fluxes at each instance of the model.  This removed the need for fitting over the luminosities, but sacrificed the ability to extract absolute luminosities directly from the results. 

As both components in the system are well outside the supported Castelli-Kurucz \citep{Castelli} grids within {\tt PHOEBE~2}, we used blackbody atmospheres along with the Johnson passband set ({\bf $V$} for the $c_2$ light curve, {\bf $B$} for He\,{\sc ii} $\lambda$\,4686, and respective bands for all \citet{Mayer08} archival datasets).  Regarding the limb-darkening model, we adopted the power model with coefficients from \citet{Claret} for both the passband and bolometric limb-darkening.  We also assumed that all incident bolometric light is reprocessed through irradiation.

For our {\tt PHOEBE~2} model, we used the two Bochum light curves, six \citet{Mayer08} archival light curves, the TESS light curve, and all RVs presented in Sect.\,\ref{omegadot}. We removed any systematics from the TESS light curve using {\tt DIPS} \citep{Prsa19}.  {\tt DIPS} separates any periodic signal at a given period from all other signals.  Here we used the known orbital period to extract the phased periodic signal from TESS, which we then unphased and placed at the mid-time of the TESS observations.  This assumes that the effects of apsidal motion are negligible during the relatively short few orbital cycles covered by the TESS light curve, but still allows accounting for the apsidal motion between the older light curves and much more recent TESS light curves, placing a stronger constraint on $\dot{\omega}$. In order to consistently handle photometric uncertainties during the MCMC analysis, we adopted per-dataset sigmas from the residuals between each individual lightcurve and the initial model.

The stellar effective temperatures of both stars were fixed to 34\,000\,K as derived from our {\tt CMFGEN} analysis (Sect.\,\ref{fitcmfgen}), the mass ratio $q$ was fixed to 1.0, and $a\sin i$ was fixed to the value of $49\,\text{R}_\odot$ derived from the RV analysis (see the last column of Table\,\ref{bestfitTable}). We assumed that stellar rotation and 
orbital motion are synchronised. For each RV dataset, we used the apparent systemic velocities per-component as determined in Sect.\,\ref{omegadot}. As preliminary values for the time-independent parameters we adopted $P_\text{orb} = 5.816475\,\text{d}$ and $e=0.134$ derived from the RV analysis (see Table\,\ref{bestfitTable}), equivalent radii $R_\text{equiv}$ of $14.2\,\text{R}_\odot$, $i=68.6^{\circ}$, and the time of primary eclipse $t_{0,\text{supconj}} = 2\,450\,549.92$  from the {\tt Nightfall} analysis (see Sect.\,\ref{photom}).

{\tt PHOEBE~2} defines all time-dependent parameters at a provided time $t_0$. In our analysis, we set $t_0 =  2\,450\,545$, the value corresponding to the mid-time of the $c_2$ light curve. The preliminary value of $\omega_0$ was set to the one determined from the {\tt Nightfall} analysis (see Table\,\ref{parambestfit}), and the preliminary value of $\dot{\omega}$ was set to the one estimated from the RV analysis (see Sect.\,\ref{omegadot}).  {\tt PHOEBE~2} then accounted for apsidal motion by adjusting $\omega(t)$ and precessing the Keplerian orbit in time through
\begin{equation}
\omega(t) = \omega_0 + \dot{\omega} (t-t_0).
\end{equation} 
While we allowed $\omega_0$ and $\dot{\omega}$ to be explored during the MCMC run, we kept $t_0$ fixed to avoid meaningless degenerate solutions between $t_0$ and $\omega_0$. $t_{0,\text{supconj}}$, which enforces the positions of the stars in the orbit to be at superior conjunction at a particular time-stamp, was allowed to vary during the MCMC solution to allow freedom in the model to shift to fit the exact timings of the eclipses.

The MCMC samples the local parameter space with a number of walkers, with each initially drawn from a provided distribution for each of the marginalised parameters.  For each iteration, the parameter values were adopted, the forward model was computed, and the $\chi^2$ between the model and the observations was determined. Each chain then decided how to move with respect to the other chains for the following iteration.  After a burn-in period during which the chains converged to the local minimum, the chains then sampled the local optimum solution in a way that allowed us to determine posteriors and uncertainties that account for the correlations and degeneracies between the marginalised parameters.

\begin{figure}[htb]
\begin{center}
\includegraphics[width=\linewidth]{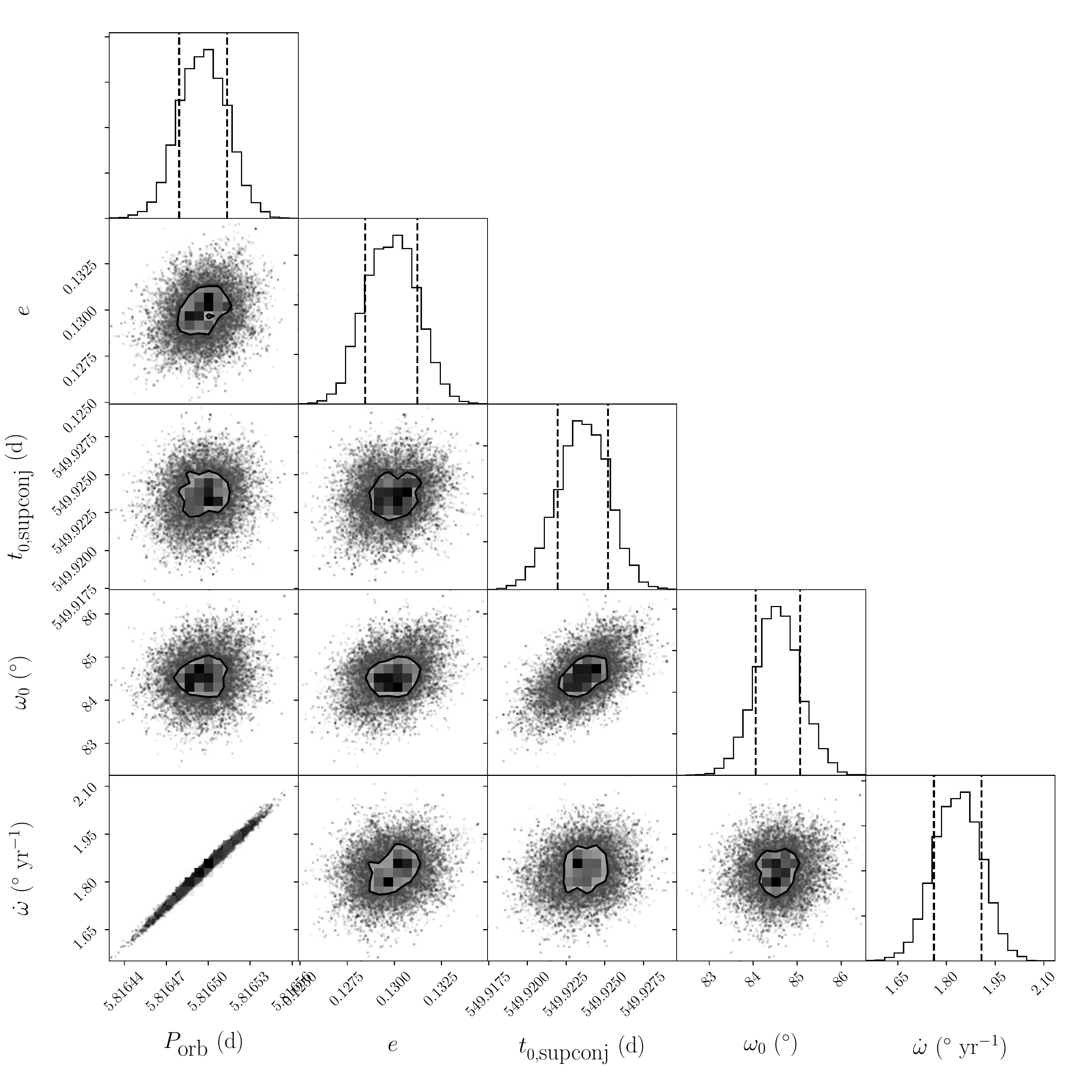}
\end{center}
\caption{Corner plot depicting the posteriors and correlations between parameters in the {\tt PHOEBE~2} analysis using the MCMC package {\tt emcee} \citep{foreman-mackey} and {\tt corner} \citep{corner}.  Only parameters related to apsidal motion are shown here. See the text for details on all marginalised parameters. } 
\label{fig:phoebe_corner}
\end{figure}

Using 72 walkers, we then allowed the local parameter space to be explored with the following free parameters: $P_\text{orb}$, $e$, $i$, $R_\text{equiv}$, $t_{0,\text{supconj}}$, $\omega_0$ , and $\dot\omega$.  Each of these were initially drawn from a small window around the preliminary values adopted, and wider uninformative priors were used to discourage any walkers from wandering too far from the solution.  After removing a burn-in of 150 iterations, the walkers randomly explored the local parameter space of the final solution.  The resulting posteriors taken from these remaining ${\sim}25\,000$ evaluations of the model are depicted in a corner plot shown in Fig.\,\ref{fig:phoebe_corner}.  By allowing the walkers to explore the parameter space in these dimensions, any uncertainties in correlated parameters are considered (and any uncertainties in fixed or assumed parameters are effectively ignored).  In Fig.\,\ref{omegaphot} we show the linear apsidal motion from these posteriors compared to the independently measured estimates of $\omega$ for each independent dataset.

We then determined the maximum likelihood  and $1\sigma$ uncertainties adopted from the 68\% confidence intervals from the 1D posteriors using {\tt chainconsumer} \citep{Hinton}, which are summarised in Table\,\ref{table:phoebe_posteriors}. These uncertainties only consider the effects and correlations due to the marginalised parameters and assume that all other values are correct at their fixed values. \\

\begin{table}[h!]
\centering
\caption{Posteriors for all marginalised parameters from the {\tt PHOEBE~2} analysis using {\tt emcee}.}
\label{table:phoebe_posteriors}
\begin{tabular}{ll}
\hline
\hline
Parameter & Posterior \\ 
\hline
\vspace*{-3mm}\\
$P_\text{orb}$ $(\text{d})$ & $5.816498^{+0.000016}_{-0.000018}$ \\ 
\vspace*{-3mm}\\
$e$ & {\bf $0.130^{+0.002}_{-0.002}$} \\ 
\vspace*{-3mm}\\
$i$ $(^\circ)$ & {\bf $67.6^{+0.2}_{-0.1}$} \\ 
\vspace*{-3mm}\\
$R_\text{equiv}$ $(\text{R}_\odot)$ & {\bf $15.07^{+0.08}_{-0.12}$} \\ 
\vspace*{-3mm}\\
$t_{0,\text{supconj}}$ $(\text{HJD})$ & {\bf $2\,450\,549.923^{+0.002}_{-0.001}$} \\ 
\vspace*{-3mm}\\
$\omega_0$ $(^\circ)$ & {\bf $84.5^{+0.5}_{-0.5}$} \\ 
\vspace*{-3mm}\\
$\dot{\omega}$ $(^\circ\,\text{yr}^{-1})$ & $1.843^{+0.064}_{-0.083}$ \\ 
\vspace*{-3mm}\\
\hline
\end{tabular}
\end{table}

The combination of our best-fit orbital inclination with the minimum semimajor axis inferred from the RV curves (see Sect.\,\ref{omegadot}) yields a semimajor axis $a$ of $53.0^{+0.3}_{-0.2}$\,R$_{\odot}$ from which individual masses of $29.5^{+0.5}_{-0.4}$\,M$_{\odot}$ were inferred for both stars.

From these values, we infer photometric values of the surface gravities for the primary and secondary stars: $\log g_1 = \log g_2 =  3.55\pm0.01$. These values are higher than those derived from spectroscopy (see discussion in Sect.\,\ref{fitcmfgen}). Using the {\tt PHOEBE} radii and the temperatures determined in Sect.\,\ref{fitcmfgen}, we infer bolometric luminosities of $(2.73\pm 0.32) \times 10^5\,L_\odot$ for both stars. These values are slightly higher than those determined from the distance in Sect.\,\ref{fitcmfgen}, but they still agree within the error bars. Table\,\ref{photometric_parameters} lists the physical parameters expressed in so-called \textit{\textup{nominal}} solar units as adopted by the International Astronomical Union \citep{NomUnit}. We note that the conventional solar units used throughout our paper agree with these nominal values. While our results indicate slightly larger stars than found by \citet{Sana01}, but slightly less massive and smaller stars than found by \citet{Nesslinger06} and \citet{Mayer08}, the parameters overlap within their error bars. Our photometric radius agrees with our spectroscopic radius determined with {\tt CMFGEN} (see Sect.\,\ref{fitcmfgen}).

\begin{table}
\caption{Physical parameters of the components of HD~152\,248 as derived from the {\tt PHOEBE} analysis.\label{photometric_parameters}}
\begin{center}
\begin{tabular}{l c c}
\hline\hline
Parameter & \multicolumn{2}{c}{Value} \\
& Primary & Secondary \\
\hline
\vspace*{-3mm}\\
$\text{M}_\text{photo}\,(\Mnom)$ & {\bf $29.5^{+0.5}_{-0.4}$} & {\bf $29.5^{+0.5}_{-0.4}$} \\
\vspace*{-3mm}\\
$R_{\rm equiv}$ (\Rnom) & $15.07^{+0.08}_{-0.12}$ & $15.07^{+0.08}_{-0.12}$ \\
$a\,(\Rnom)$ & \multicolumn{2}{c}{{\bf $53.0^{+0.3}_{-0.2}$}} \\
\vspace*{-3mm}\\
$\log g_\text{photo}$\,(cgs) & $3.55\pm0.01$ & $3.55\pm0.01$ \\
\vspace*{-3mm}\\
$L_\text{bol,photo}\,(\Lnom)$ & $(2.73\pm 0.32) \times 10^5$ & $(2.73\pm 0.32) \times 10^5$ \\
\hline
\end{tabular}
\end{center}
\end{table}

Assuming the rotational axes of the stars are orthogonal to the orbital plane, we can combine the best-fit values of the inclination and stellar radii with the projected rotational velocities of the stars determined in Sect.\,\ref{sect:vsini} to derive rotational periods of $5.11\pm0.15$\,days for the primary and $5.14\pm0.23$\,days for the secondary. The stars have their rotation synchronised with each other to within the error bars. We compare the rotational angular velocities with the instantaneous orbital angular velocity at periastron and find ratios of $0.87\pm0.03$ for the primary and $0.86\pm0.04$ for the secondary. We conclude that pseudo-synchronisation is about to be achieved in the system for both stars.

\section{Internal structure constant $k_2$ \label{sect:k2}}
The rate of apsidal motion can be expressed as the sum of a classical Newtonian contribution (N) and a general relativistic correction (GR), 
\begin{equation}
\dot\omega = \dot\omega_\text{N} + \dot\omega_\text{GR}.
\end{equation}

Assuming that the rotation axes of the stars are aligned with the normal to the orbital plane, the Newtonian term takes the form adopted by \citet{sterne}, where only the contributions arising from the second-order harmonic distortions of the potential are considered:
\begin{equation}
\label{eqn:omegadotN}
\begin{aligned}
\dot\omega_\text{N} = \frac{2\pi}{P_\text{orb}} \Bigg[&15f(e)\left\{k_{2,1}q \left(\frac{R_1}{a}\right)^5 + \frac{k_{2,2}}{q} \left(\frac{R_2}{a}\right)^5\right\} \\
& \begin{aligned} + g(e) \Bigg\{ &k_{2,1} (1+q) \left(\frac{R_1}{a}\right)^5 \left(\frac{P_\text{orb}}{P_\text{rot,1}}\right)^2 \\ 
& + k_{2,2}\, \frac{1+q}{q} \left(\frac{R_2}{a}\right)^5 \left(\frac{P_\text{orb}}{P_\text{rot,2}}\right)^2 \Bigg\}  \Bigg],
\end{aligned}
\end{aligned}
\end{equation}
where $a$ is the semimajor axis, $q=m_2/m_1$ is the mass ratio, $R_1$ and $R_2$ are the radii of primary and secondary star, respectively, $k_{2,1}$ and $k_{2,2}$ are the apsidal motion constants of primary and secondary star, respectively, $P_\text{rot,1}$ and $P_\text{rot,2}$ are the rotational periods of primary and secondary star, respectively, and where the functions $f$ and $g$ depend only on the eccentricity through the following relations:
\begin{equation}
  \label{eqn:fg}
\left\{
\begin{aligned}
& f(e) = \frac{1+\frac{3e^2}{2}+\frac{e^4}{8}}{(1-e^2)^5}, \\ 
  & g(e) = \frac{1}{(1-e^2)^2}.
\end{aligned}
\right. 
\end{equation}
The Newtonian term is itself the sum of the effects induced by tidal deformation on the one hand and rotation of the stars on the other hand.

The general relativistic contribution to the rate of apsidal motion, when only the quadratic corrections are taken into account, is given by the expression 
\begin{equation}
\label{eqn:omegadotGR}
\begin{aligned}
\dot\omega_\text{GR} &= \frac{2\pi}{P_\text{orb}}\frac{3G(m_1+m_2)}{c^2 a (1-e^2)} \\
& = \left(\frac{2\pi}{P_\text{orb}}\right)^{5/3}\frac{3(G(m_1+m_2))^{2/3}}{c^2 (1-e^2)},
\end{aligned}
\end{equation} 
where $G$ is the gravitational constant and $c$ is the speed of light \citep{Shakura}. 

Using the orbital period of the system, its eccentricity and the masses of the stars determined from the observations, we infer a value of  $(0.163\pm 0.001)^{\circ}\,\text{yr}^{-1}$ for the general relativistic contribution to the total observational rate of apsidal motion $\dot\omega_{\text{GR,obs}}$. We then deduce the Newtonian contribution: $\dot\omega_{\text{N,obs}}=(1.680^{+0.064}_{-0.083})^{\circ}\,\text{yr}^{-1}$. 

We can compute the observational internal structure constants $k_{2,\text{obs}}$ using Eqs.\,\ref{eqn:omegadotN} and \ref{eqn:fg} with the value inferred for $\dot\omega_{N,\text{obs}}$ and taking both radii and masses of the stars obtained with the {\tt PHOEBE} analysis. To solve for $k_{2,\text{obs}}$, we have to assume that both stars have the same internal structure constants. Because the two stars are nearly exactly identical, this assumption appears reasonable. In this way, we find   $k_{2,\text{obs}}=0.0010\pm 0.0001$.

\section{Conclusion \label{conclusion}}
We have reanalysed the eccentric massive binary HD~152\,248, the most emblematic eclipsing O-star binary in the open cluster NGC~6231. We reconstructed the individual spectra of the components using a disentangling code. The analysis of these spectra by means of the {\tt CMFGEN} model atmosphere code further allowed us to determine stellar and wind properties of the system. Orbital inclination and Roche-lobe filling factors were constrained through the analysis of the light curve by means of the {\tt Nightfall} code. From the {\tt PHOEBE} analysis, mean stellar radii of $15.07^{+0.08}_{-0.12}~\Rnom$ and masses of $29.5^{+0.5}_{-0.4}~\Mnom$ were obtained for both stars. 
 
Radial velocity data from literature and redetermined in this paper as well as photometric data allowed us to establish a rate of apsidal motion of  $(1.843^{+0.064}_{-0.083})^{\circ}$\,yr$^{-1}$ by means of the {\tt PHOEBE} code, which corresponds to a period of $195^{+7}_{-9}$\,yr. The internal structure constant $k_2$ was then observationally constrained to a value of $0.0010\pm 0.0001$ for both stars. 

In contrast to what has been envisaged by some authors \citep{Penny,Sana01}, there is no reason to consider that the evolutionary states of the components of HD~152\,248 were influenced by an ongoing or past Roche-lobe overflow episode. The stars are currently far from filling their Roche lobes, even around periastron passage. Moreover, our spectroscopic analysis revealed no indication of a past episode of mass and angular momentum transfer. In contrast to systems that have gone through such a mass exchange episode \citep[e.g.,\ HD~47\,129, HD~149\,404, and LSS~3074;][]{Linder,Raucq,Raucq17}, the CNO surface abundances only slightly deviate from solar and are essentially identical in both stars, and there is no strong asynchronism in the rotational velocities of both components.

Each of the methods employed to establish the rate of apsidal motion and the fundamental parameters of the stars has its own systematic errors. We therefore combined various complementary approaches here to achieve results that are as robust as possible.

The properties of HD~152\,248 render the system an interesting target to study tidally induced apsidal motion from a theoretical point of view. As a next step, we will use stellar evolution models to determine theoretical rates of apsidal motion and compare them to the observed rate (Rosu et al. 2020, in preparation). This should allow identifying the most important mixing processes inside the stars as well as inferring an age estimate for the binary system.

\begin{acknowledgements}
The Li\`ege team acknowledges support from the Fonds de la Recherche Scientifique (F.R.S.- FNRS, Belgium). We thank Drs John Hillier and Rainer Wichmann for making their codes, respectively {\tt CMFGEN} and {\tt Nightfall}, publicly available. This paper makes use of data collected by the TESS mission, whose funding is provided by the NASA Explorer Program. 
\end{acknowledgements}

\begin{appendix} 

\section{Journal of the spectroscopic observations\label{appendix:spectrotable}}
This appendix provides the journal of spectroscopic observations of HD~152\,248 (Table\,\ref{Table:spectro+RV}).  
\begin{table*}[htb]
\caption{Journal of the spectroscopic observations of HD~152\,248. Column\,1 gives the heliocentric Julian date (HJD) of the observations at mid-exposure. Column\,2 gives the observational phase $\phi$ computed with the orbital period determined in Sect.\,\ref{omegadot} (last column of Table\,\ref{bestfitTable}). Columns\,3 and 4 give the radial velocity $RV_1$ and $RV_2$ of primary and secondary star, respectively. The errors represent $\pm1\sigma$. Column\,5 provides information about the instrumentation.     }
\begin{center}
\begin{tabular}{c c r r l}
\hline\hline
HJD & $\phi$  & $RV_1$ & $RV_2$ & Instrumentation\\
--\,2\,450\,000&  & (km\,s$^{-1}$) & (km\,s$^{-1}$) & \\
\hline
1299.817 & 0.941 & $88.2\pm2.7$ & $-133.2\pm1.4$ & ESO 1.5 m + FEROS\\
1300.809 & 0.111 & $-181.1\pm0.9$ & $134.2\pm1.4$ & ESO 1.5 m + FEROS\\
1301.815 & 0.284 & $-227.7\pm1.4$ & $174.9\pm1.0$ & ESO 1.5 m + FEROS\\
1304.886 & 0.812 & $193.1\pm1.2$ & $-236.1\pm0.8$ &ESO 1.5 m + FEROS \\
1304.891 & 0.813 & $191.3\pm1.1$ & $-236.2\pm0.7$ &ESO 1.5 m + FEROS \\
1327.780 & 0.748 & $177.3\pm1.2$ & $-225.0\pm0.6$ & ESO 1.5 m + FEROS\\
1668.828 & 0.383 & $-151.1\pm2.7$ & $86.0\pm2.6$ & ESO 1.5 m + FEROS\\
1668.836 & 0.384 & $-151.0\pm2.4$ & $83.9\pm2.3$ &ESO 1.5 m + FEROS \\
1670.832 & 0.728 & $177.1\pm1.2$ & $-211.7\pm0.8$ & ESO 1.5 m + FEROS\\
1670.838 & 0.729 & $177.0\pm1.3$ & $-212.6\pm0.9$ & ESO 1.5 m + FEROS\\
1671.836 & 0.900 & $138.5\pm1.1$ & $-187.3\pm1.2$ & ESO 1.5 m + FEROS\\
1671.842 & 0.901 & $138.5\pm1.7$ & $-186.4\pm1.4$ & ESO 1.5 m + FEROS\\
1672.825 & 0.070 & $-141.2\pm0.9$ & $77.1\pm1.4$ & ESO 1.5 m + FEROS\\
1672.831 & 0.071 & $-142.6\pm1.0$ & $79.0\pm1.3$ & ESO 1.5 m + FEROS\\
1673.801 & 0.238 & $-232.8\pm0.8$ & $183.0\pm1.2$ & ESO 1.5 m + FEROS\\
1673.809 & 0.239 & $-233.2\pm0.8$ & $183.8\pm1.2$ & ESO 1.5 m + FEROS\\
1673.817 & 0.241 & $-232.7\pm0.8$ & $183.9\pm1.3$ & ESO 1.5 m + FEROS\\
3546.769 & 0.248 & $-221.2\pm0.7$ & $182.2\pm1.7$ & ESO 2.2 m + FEROS\\
5642.820 & 0.614 & $133.5\pm1.0$ & $-176.4\pm2.5$ & ESO 2.2 m + FEROS\\
6761.037 & 0.864 & $103.3\pm2.1$ & $-142.5\pm1.7$ & CFHT 3.6 m+ ESPaDOns\\
6761.044 & 0.865 & $103.1\pm2.3$ & $-140.7\pm1.9$ & CFHT 3.6 m+ ESPaDOns\\
6761.047 & 0.866 & $102.1\pm2.3$ & $-140.0\pm1.9$ & CFHT 3.6 m+ ESPaDOns\\
6761.050 & 0.867 & $100.9\pm2.2$ & $-139.8\pm2.0$ & CFHT 3.6 m+ ESPaDOns\\
6761.057 & 0.868 & $99.8\pm2.5$ & $-136.9\pm1.9$ & CFHT 3.6 m+ ESPaDOns\\
6761.064 & 0.869 & $97.8\pm2.6$ & $-135.2\pm1.8$ & CFHT 3.6 m+ ESPaDOns\\
6762.026 & 0.034 & $-175.3\pm1.0$ & $119.3\pm1.0$ & CFHT 3.6 m+ ESPaDOns\\
6762.033 & 0.035 & $-177.3\pm1.1$ & $121.2\pm1.0$ & CFHT 3.6 m+ ESPaDOns\\
6762.036 & 0.036 & $-178.3\pm1.0$ & $121.4\pm1.0$ & CFHT 3.6 m+ ESPaDOns\\
6762.040 & 0.037 & $-180.1\pm1.0$ & $121.9\pm1.0$ & CFHT 3.6 m+ ESPaDOns\\
6762.046 & 0.038 & $-181.5\pm1.0$ & $124.3\pm1.0$ & CFHT 3.6 m+ ESPaDOns\\
6762.053 & 0.039 & $-184.7\pm0.9$ & $124.3\pm1.0$ & CFHT 3.6 m+ ESPaDOns\\
6762.060 & 0.040 & $-185.1\pm0.9$ & $128.4\pm1.0$ & CFHT 3.6 m+ ESPaDOns\\
6762.062 & 0.041 & $-187.1\pm0.9$ & $128.9\pm1.0$ & CFHT 3.6 m+ ESPaDOns\\
6762.063 & 0.041 & $-186.1\pm0.9$ & $128.6\pm1.0$ & CFHT 3.6 m+ ESPaDOns\\
6762.073 & 0.042 & $-188.0\pm0.9$ & $131.3\pm1.0$ & CFHT 3.6 m+ ESPaDOns\\
6762.080 & 0.044 & $-190.2\pm0.9$ & $133.2\pm0.9$ & CFHT 3.6 m+ ESPaDOns\\
6762.087 & 0.045 & $-191.6\pm0.8$ & $134.2\pm1.1$ & CFHT 3.6 m+ ESPaDOns\\
6762.093 & 0.046 & $-193.5\pm0.9$ & $136.2\pm1.1$ & CFHT 3.6 m+ ESPaDOns\\
6762.090 & 0.045 & $-192.7\pm0.8$ & $135.6\pm1.0$ & CFHT 3.6 m+ ESPaDOns\\
6762.100 & 0.047 & $-194.3\pm0.9$ & $138.5\pm1.0$ & CFHT 3.6 m+ ESPaDOns\\
\hline
\end{tabular}
\end{center} 
\label{Table:spectro+RV}
\end{table*}

\section{New photometric observations of HD~152\,248 \label{appendixB}}
In this appendix, we provide two tables  with the photometric data collected in 1997 at the 0.6\,m Bochum telescope at La Silla observatory. The data were taken through two narrow-band filters, centred on the He\,{\sc ii} $\lambda$\,4686 line (Tables\,\ref{journalHeII}) and the $c_2$ continuum near 6051\AA\ (Tables\,\ref{journalc2}) \citep{Royer}.

\begin{table*}[h!]
\centering
\footnotesize
\caption{Journal of the photometric observations through the He\,{\sc ii} $\lambda$\,4686 filter. }
\label{journalHeII}
\begin{minipage}{8cm}
\begin{tabular}{ccc}
\hline\hline
HJD-2\,450\,000 & Differential magnitude & Error\\
& (mag) & (mag) \\
\hline
534.8722 & 0.0008 & 0.010 \\
535.7976 & 0.0145 & 0.010 \\
535.8262 & 0.0208 & 0.010 \\
535.9103 & 0.0078 & 0.010 \\
537.7754 & $-$0.0087 & 0.010 \\
537.8168 & $-$0.0030 & 0.010 \\
537.8576 & 0.0098 & 0.010 \\
537.8590 & 0.0053 & 0.010 \\
538.7595 & $-$0.0095 & 0.010 \\
538.9045 & 0.0040 & 0.010 \\
539.7730 & $-$0.0155 & 0.010 \\
539.8246 & $-$0.0190 & 0.010 \\
539.8655 & 0.0133 & 0.010 \\
539.8983 & $-$0.0225 & 0.010 \\
540.7890 & 0.0068 & 0.010 \\
540.8306 & 0.0083 & 0.010 \\
540.8815 & 0.0235 & 0.010 \\
540.9135 & 0.0232 & 0.010 \\
540.9217 & 0.0303 & 0.010 \\
540.9280 & 0.0283 & 0.010 \\
541.7689 & 0.0028 & 0.010 \\
541.7931 & 0.0130 & 0.010 \\
541.8423 & 0.0085 & 0.010 \\
541.8681 & 0.0070 & 0.010 \\
542.7503 & $-$0.0150 & 0.010 \\
542.7888 & $-$0.0180 & 0.010 \\
542.8183 & $-$0.0090 & 0.010 \\
542.8474 & $-$0.0237 & 0.010 \\
542.8898 & $-$0.0205 & 0.010 \\
543.7516 & 0.0220 & 0.010 \\
543.8076 & 0.0430 & 0.010 \\
543.8693 & 0.0888 & 0.010 \\
543.8947 & 0.1085 & 0.010 \\
543.9218 & 0.1173 & 0.010 \\
544.7840 & $-$0.0182 & 0.010 \\
544.8465 & $-$0.0285 & 0.010 \\
544.8928 & $-$0.0225 & 0.010 \\
544.9202 & $-$0.0215 & 0.010 \\
544.9272 & $-$0.0210 & 0.010 \\
545.7744 & $-$0.0227 & 0.010 \\
545.8333 & $-$0.0295 & 0.010 \\
545.8758 & $-$0.0145 & 0.010 \\
545.9226 & $-$0.0050 & 0.010 \\
546.7415 & 0.0245 & 0.010 \\
546.7979 & 0.0355 & 0.010 \\
546.8556 & 0.0483 & 0.010 \\
546.9197 & 0.0703 & 0.010 \\
547.7870 & 0.0025 & 0.010 \\
547.8468 & $-$0.0030 & 0.010 \\
547.8905 & $-$0.0140 & 0.010 \\
548.7645 & $-$0.0247 & 0.010 \\
548.8095 & $-$0.0305 & 0.010 \\
\hline
\end{tabular}
\end{minipage}
\hfill
\begin{minipage}{8cm}
\begin{tabular}{ccc}
\hline\hline
HJD-2\,450\,000 & Differential magnitude & Error\\
& (mag) & (mag) \\
\hline
548.8509 & $-$0.0275 & 0.010 \\
548.8781 & $-$0.0385 & 0.010 \\
548.9168 & $-$0.0335 & 0.010 \\
548.9212 & $-$0.0270 & 0.010 \\
549.7466 & 0.1448 & 0.010 \\
549.8067 & 0.1648 & 0.010 \\
549.8637 & 0.1830 & 0.010 \\
549.9214 & 0.1860 & 0.010 \\
550.7748 & $-$0.0362 & 0.010 \\
550.8190 & $-$0.0230 & 0.010 \\
550.8586 & $-$0.0495 & 0.010 \\
550.8864 & $-$0.0362 & 0.010 \\
550.9155 & $-$0.0270 & 0.010 \\
550.9208 & $-$0.0072 & 0.010 \\
551.7489 & $-$0.0095 & 0.010 \\
551.7930 & $-$0.0102 & 0.010 \\
551.8052 & $-$0.0102 & 0.010 \\
551.8623 & $-$0.0030 & 0.010 \\
551.9173 & $-$0.0102 & 0.010 \\
552.7569 & 0.0533 & 0.010 \\
552.7614 & 0.0563 & 0.010 \\
552.8551 & 0.0755 & 0.010 \\
552.8782 & 0.0758 & 0.010 \\
552.9174 & 0.0773 & 0.010 \\
553.7500 & 0.0000 & 0.010 \\
553.7941 & 0.0065 & 0.010 \\
553.8321 & $-$0.0062 & 0.010 \\
553.8729 & $-$0.0082 & 0.010 \\
553.9188 & $-$0.0067 & 0.010 \\
554.7062 & $-$0.0337 & 0.010 \\
554.7533 & $-$0.0420 & 0.010 \\
554.7928 & $-$0.0387 & 0.010 \\
554.8312 & $-$0.0162 & 0.010 \\
554.8680 & $-$0.0285 & 0.010 \\
554.9149 & $-$0.0327 & 0.010 \\
555.7662 & 0.1950 & 0.010 \\
555.8110 & 0.1840 & 0.010 \\
555.8573 & 0.1703 & 0.010 \\
555.9148 & 0.1195 & 0.010 \\
555.9303 & 0.1160 & 0.010 \\
556.7881 & $-$0.0280 & 0.010 \\
556.8690 & $-$0.0330 & 0.010 \\
557.7451 & 0.0043 & 0.010 \\
557.7899 & $-$0.0035 & 0.010 \\
557.8334 & 0.0078 & 0.010 \\
557.8599 & 0.0030 & 0.010 \\
557.9249 & 0.0015 & 0.010 \\
558.6961 & 0.0780 & 0.010 \\
558.7425 & 0.0998 & 0.010 \\
558.7897 & 0.0903 & 0.010 \\
558.8082 & 0.0890 & 0.010 \\
& & \\
\hline
\end{tabular}
\end{minipage}
\tablefoot{HJD correspond to dates at mid-exposure.}
\end{table*}

\begin{table*}[h!]
\centering
\footnotesize
\caption{Journal of the photometric observations through the $c_2$ filter. }
\label{journalc2}
\begin{minipage}{8cm}
\begin{tabular}{ccc}
\hline\hline
HJD-2\,450\,000 & Differential magnitude & Error\\
& (mag) & (mag) \\
\hline
530.8192 & $-$0.0063 & 0.010 \\
531.8341 & $-$0.0145 & 0.010 \\
531.8354 & $-$0.0192 & 0.010 \\
533.8033 & $-$0.0152 & 0.010 \\
533.8041 & $-$0.0155 & 0.010 \\
533.8291 & $-$0.0173 & 0.010 \\
533.8299 & $-$0.0085 & 0.010 \\
533.8307 & $-$0.0135 & 0.010 \\
533.8571 & $-$0.0120 & 0.010 \\
533.8579 & $-$0.0135 & 0.010 \\
533.8587 & $-$0.0168 & 0.010 \\
533.8823 & $-$0.0122 & 0.010 \\
533.8831 & $-$0.0090 & 0.010 \\
533.8839 & $-$0.0128 & 0.010 \\
533.9053 & $-$0.0140 & 0.010 \\
533.9063 & $-$0.0128 & 0.010 \\
533.9270 & $-$0.0023 & 0.010 \\
533.9278 & $-$0.0017 & 0.010 \\
533.9286 & $-$0.0115 & 0.010 \\
533.9294 & $-$0.0088 & 0.010 \\
533.9302 & $-$0.0072 & 0.010 \\
534.7974 & $-$0.0068 & 0.010 \\
534.8227 & $-$0.0060 & 0.010 \\
534.8456 & $-$0.0042 & 0.010 \\
534.8730 & $-$0.0042 & 0.010 \\
534.8976 & 0.0012 & 0.010 \\
535.7984 & 0.0245 & 0.010 \\
535.8270 & 0.0242 & 0.010 \\
535.8678 & 0.0217 & 0.010 \\
535.9111 & 0.0185 & 0.010 \\
537.7762 & 0.0020 & 0.010 \\
537.8176 & 0.0045 & 0.010 \\
537.9021 & 0.0250 & 0.010 \\
538.7603 & $-$0.0065 & 0.010 \\
538.8030 & $-$0.0032 & 0.010 \\
538.8623 & $-$0.0115 & 0.010 \\
538.9053 & 0.0027 & 0.010 \\
539.7738 & $-$0.0090 & 0.010 \\
539.8254 & $-$0.0020 & 0.010 \\
539.8663 & 0.0328 & 0.010 \\
539.8991 & $-$0.0125 & 0.010 \\
540.7898 & 0.0128 & 0.010 \\
540.8314 & 0.0245 & 0.010 \\
540.8836 & 0.0052 & 0.010 \\
540.9142 & 0.0195 & 0.010 \\
540.9223 & 0.0320 & 0.010 \\
541.7697 & 0.0055 & 0.010 \\
541.7939 & 0.0050 & 0.010 \\
541.8431 & 0.0072 & 0.010 \\
541.8689 & $-$0.0023 & 0.010 \\
541.8974 & 0.0072 & 0.010 \\
542.7511 & $-$0.0155 & 0.010 \\
542.7896 & $-$0.0162 & 0.010 \\
542.8482 & $-$0.0140 & 0.010 \\
542.8906 & $-$0.0125 & 0.010 \\
543.7525 & 0.0145 & 0.010 \\
543.8084 & 0.0412 & 0.010 \\
543.8701 & 0.0862 & 0.010 \\
543.8955 & 0.1075 & 0.010 \\
543.9226 & 0.1202 & 0.010 \\
544.7848 & $-$0.0128 & 0.010 \\
544.8936 & $-$0.0210 & 0.010 \\
544.9210 & $-$0.0238 & 0.010 \\
\hline
\end{tabular}
\end{minipage}
\hfill
\begin{minipage}{8cm}
\begin{tabular}{ccc}
\hline\hline
HJD-2\,450\,000 & Differential magnitude & Error\\
& (mag) & (mag) \\
\hline
544.9281 &0.0015 & 0.010 \\
545.7752 & $-$0.0230 & 0.010 \\
545.8341 & $-$0.0137 & 0.010 \\
545.8766 & $-$0.0057 & 0.010 \\
545.9234 & $-$0.0112 & 0.010 \\
546.7423 & 0.0255 & 0.010 \\
546.7987 & 0.0440 & 0.010 \\
546.8564 & 0.0537 & 0.010 \\
546.9205 & 0.0745 & 0.010 \\
547.7881 & 0.0042 & 0.010 \\
547.8476 & 0.0010 & 0.010 \\
547.8913 & $-$0.0045 & 0.010 \\
547.9250 & 0.0023 & 0.010 \\
548.7674 & $-$0.0213 & 0.010 \\
548.8105 & $-$0.0322 & 0.010 \\
548.8519 & $-$0.0300 & 0.010 \\
548.8789 & $-$0.0345 & 0.010 \\
548.9179 & $-$0.0268 & 0.010 \\
548.9200 & $-$0.0280 & 0.010 \\
548.9223 & $-$0.0245 & 0.010 \\
549.7474 & 0.1405 & 0.010 \\
549.8075 & 0.1660 & 0.010 \\
549.8645 & 0.1782 & 0.010 \\
549.9225 & 0.2165 & 0.010 \\
550.7756 & $-$0.0300 & 0.010 \\
550.8198 & $-$0.0045 & 0.010 \\
550.8872 & $-$0.0195 & 0.010 \\
551.7497 & 0.0040 & 0.010 \\
551.7937 & 0.0000 & 0.010 \\
551.8631 & 0.0010 & 0.010 \\
552.7605 & 0.0590 & 0.010 \\
552.8081 & 0.0673 & 0.010 \\
552.8566 & 0.0750 & 0.010 \\
552.8789 & 0.0772 & 0.010 \\
552.8798 & 0.0755 & 0.010 \\
552.9189 & 0.0800 & 0.010 \\
553.7507 & 0.0005 & 0.010 \\
553.7947 & 0.0050 & 0.010 \\
553.8328 & 0.0008 & 0.010 \\
553.8736 & $-$0.0135 & 0.010 \\
554.7069 & $-$0.0340 & 0.010 \\
554.7540 & $-$0.0402 & 0.010 \\
554.7937 & $-$0.0417 & 0.010 \\
554.8319 & $-$0.0115 & 0.010 \\
554.8686 & $-$0.0278 & 0.010 \\
554.9156 & $-$0.0312 & 0.010 \\
555.7669 & 0.1935 & 0.010 \\
555.8117 & 0.1845 & 0.010 \\
555.8581 & 0.1635 & 0.010 \\
555.9156 & 0.1178 & 0.010 \\
555.9311 & 0.1230 & 0.010 \\
556.7887 & 0.0032 & 0.010 \\
556.8698 & $-$0.0250 & 0.010 \\
557.7458 & 0.0175 & 0.010 \\
557.7907 & 0.0110 & 0.010 \\
557.8342 & 0.0135 & 0.010 \\
557.8607 & 0.0055 & 0.010 \\
557.9094 & 0.0060 & 0.010 \\
557.9257 & 0.0122 & 0.010 \\
558.6969 & 0.0873 & 0.010 \\
558.7433 & 0.0902 & 0.010 \\
558.7910 & 0.0990 & 0.010 \\
& & \\
\hline
\end{tabular}
\end{minipage}
\tablefoot{HJD correspond to dates at mid-exposure.}
\end{table*}

\section{Fourier transform of line profiles\label{appendix:fourier}}
For the lines we used to determine the projected rotational velocities (Sect.\,\ref{sect:vsini}, Table\,\ref{vsiniTable}), the corresponding line profiles of the separated spectra obtained after application of the brightness ratio for the primary and secondary stars are presented in Fig.\,\ref{Fig:vsini}\subref{subfig:SiIV}-\subref{subfig:HeI5016}. The Fourier transform of these lines as well as the best-match rotational profiles for the primary and secondary stars are also shown in this figure.

\begin{figure*}[htb]
\begin{center}
\subfigure[Si\,{\sc iv} $\lambda$\,4089 line.]{\includegraphics[width=\linewidth, trim=0.5cm 7.8cm 0cm 2cm, clip=true]{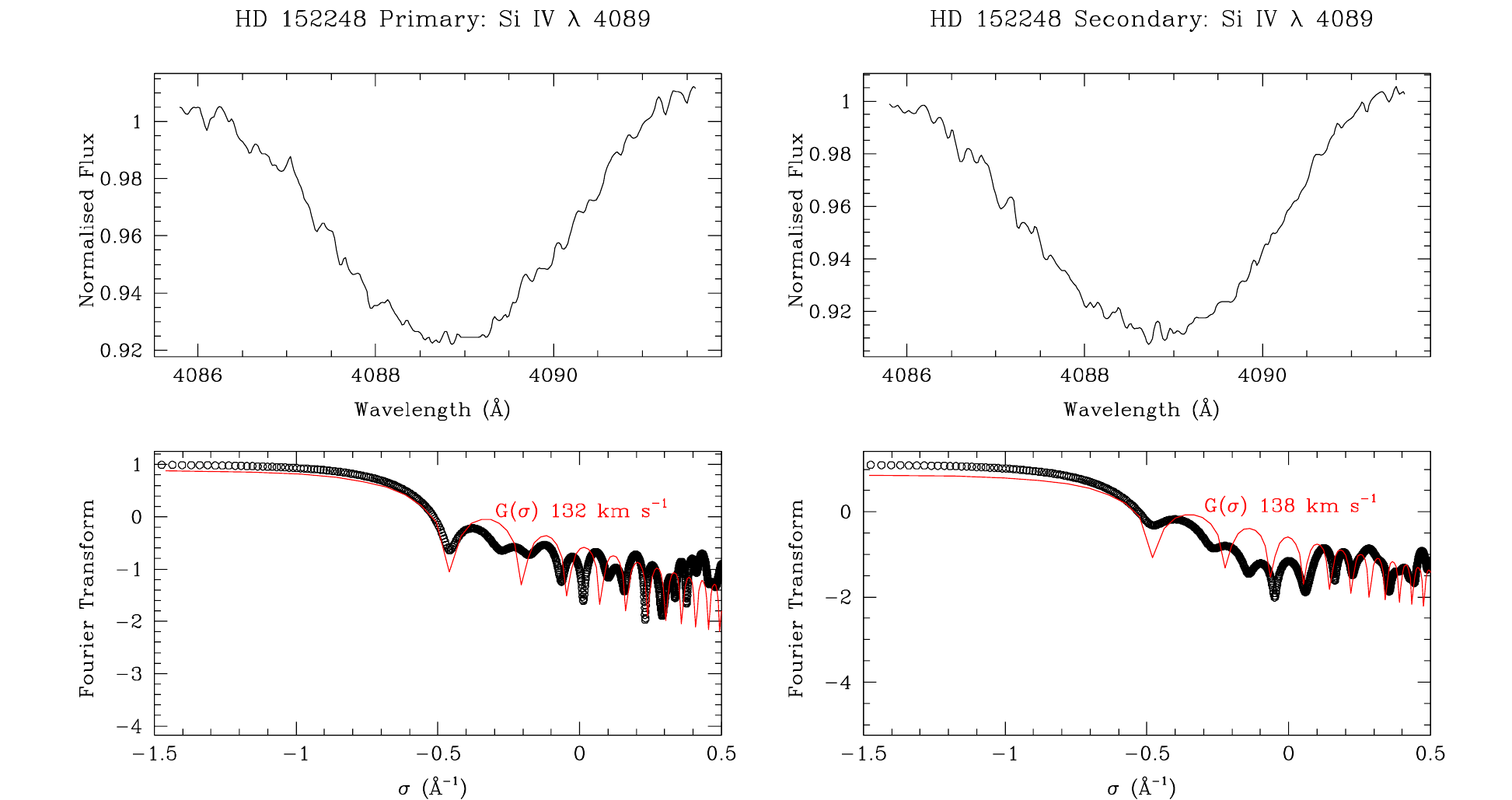}\label{subfig:SiIV}}
\subfigure[O\,{\sc iii} $\lambda$\,5592 line.]{\includegraphics[width=\linewidth, trim=0.5cm 7.8cm 0cm 2cm, clip=true]{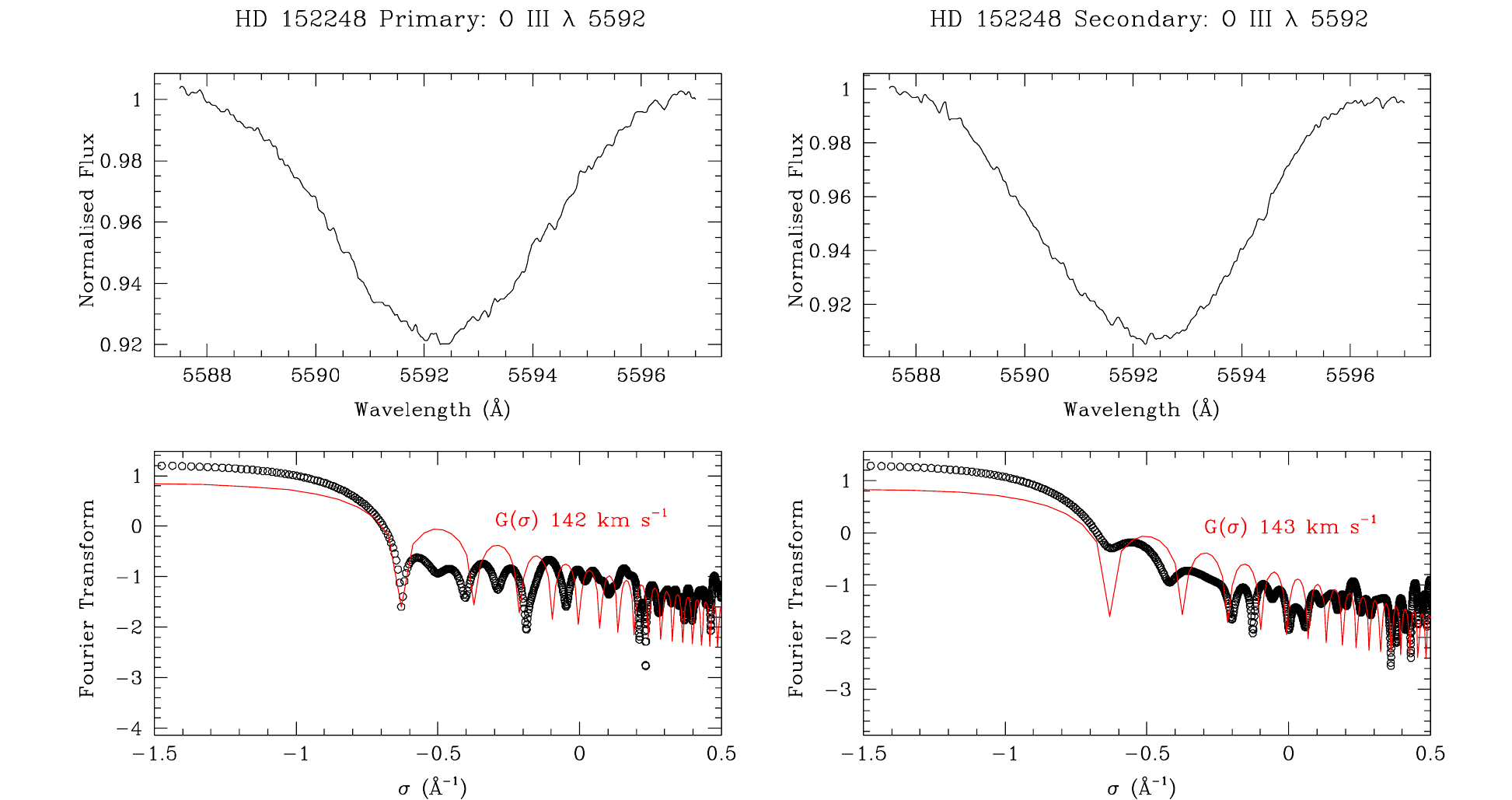}\label{subfig:OIII}}
\end{center}
\caption{\textit{Top row:} Line profiles (\subref{subfig:SiIV} Si\,{\sc iv} $\lambda$\,4089, \subref{subfig:OIII} O\,{\sc iii} $\lambda$\,5592, \subref{subfig:CIV} C\,{\sc iv} $\lambda$\,5812, \subref{subfig:HeI4120} He\,{\sc i} $\lambda$\,4120, \subref{subfig:HeI4713} He\,{\sc i} $\lambda$\,4713, \subref{subfig:HeI4922} He\,{\sc i} $\lambda$\,4922, and \subref{subfig:HeI5016} He\,{\sc i} $\lambda$\,5016) of the separated spectra obtained after application of the brightness ratio for the primary (\textit{left panel}) and secondary (\textit{right panel}) stars. \textit{Bottom row:} Fourier transform of these lines (in black) and best-match rotational profile (in red) for the primary (\textit{left panel}) and secondary (\textit{right panel}) stars. \label{Fig:vsini}}
\end{figure*}

\begin{figure*}[htb]
\ContinuedFloat
\setcounter{subfigure}{2}
\begin{center}
\subfigure[C\,{\sc iv} $\lambda$\,5812 line.]{\includegraphics[width=\linewidth, trim=0.5cm 7.8cm 0cm 2cm, clip=true]{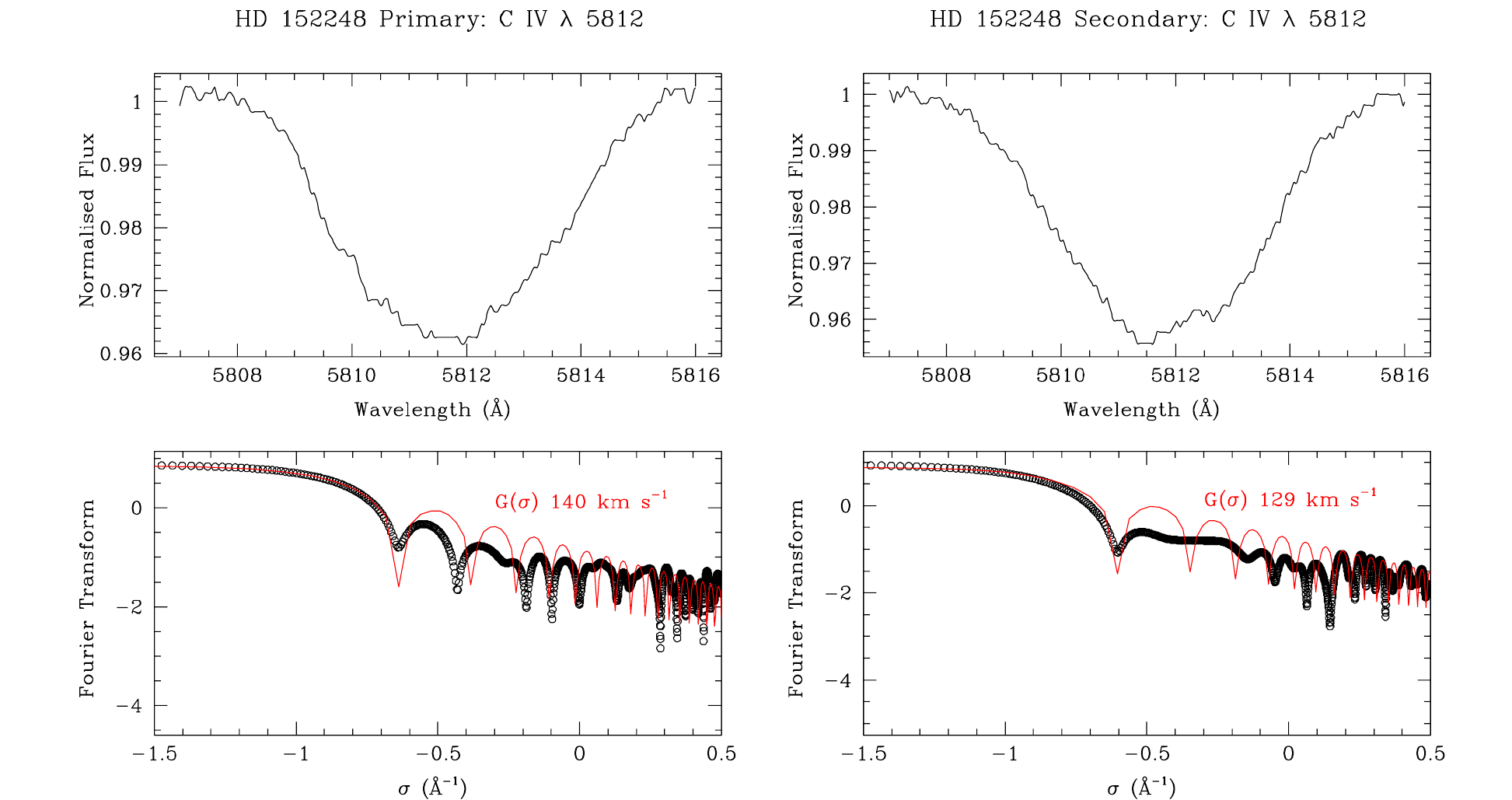}\label{subfig:CIV}}
\subfigure[He\,{\sc i} $\lambda$\,4120 line.]{\includegraphics[width=\linewidth, trim=0.5cm 7.8cm 0cm 2cm, clip=true]{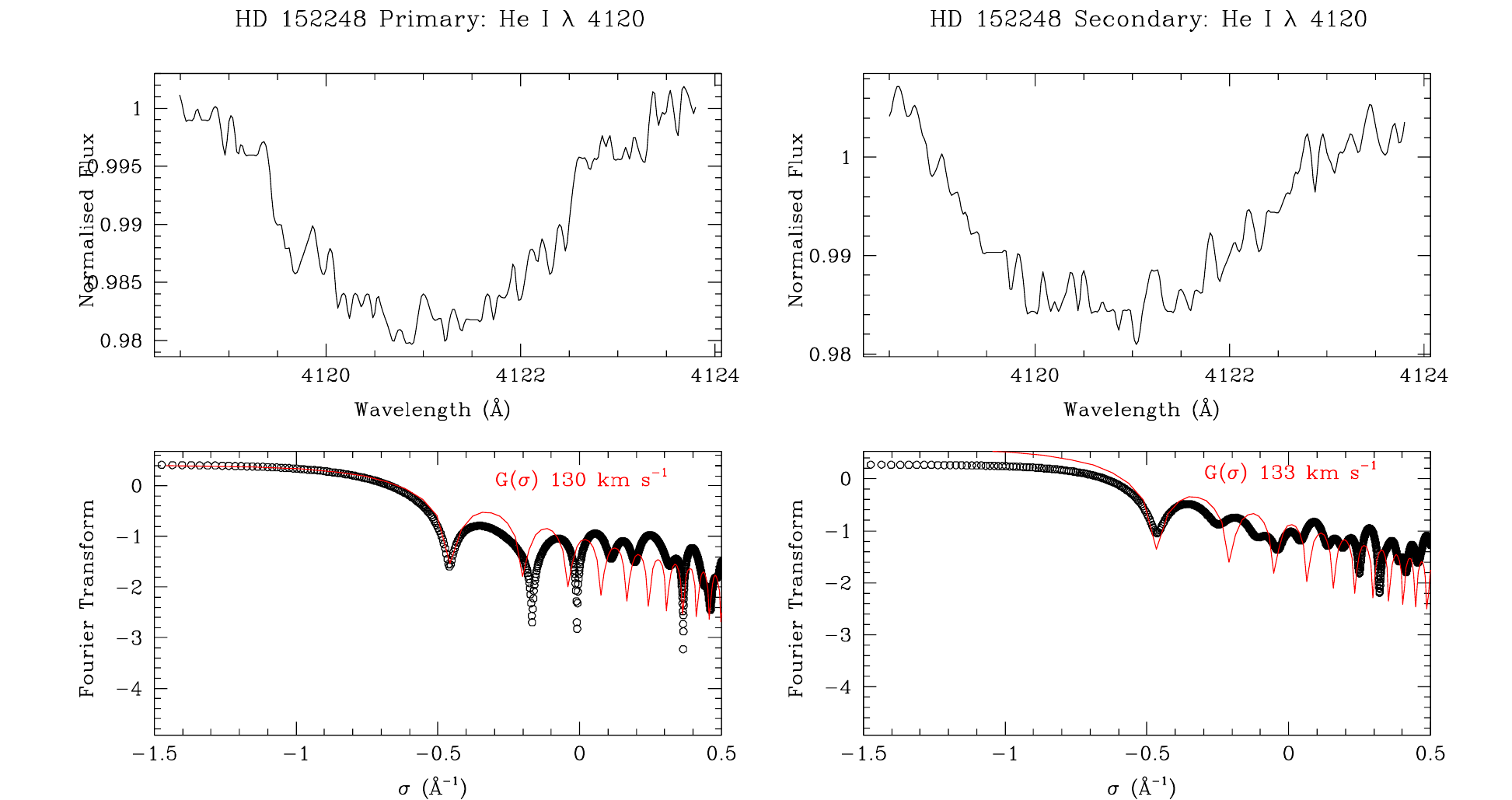}\label{subfig:HeI4120}}
\end{center}
\caption{\textit{Top row:} Line profiles (\subref{subfig:SiIV} Si\,{\sc iv} $\lambda$\,4089, \subref{subfig:OIII} O\,{\sc iii} $\lambda$\,5592, \subref{subfig:CIV} C\,{\sc iv} $\lambda$\,5812, \subref{subfig:HeI4120} He\,{\sc i} $\lambda$\,4120, \subref{subfig:HeI4713} He\,{\sc i} $\lambda$\,4713, \subref{subfig:HeI4922} He\,{\sc i} $\lambda$\,4922, and \subref{subfig:HeI5016} He\,{\sc i} $\lambda$\,5016) of the separated spectra obtained after application of the brightness ratio for the primary (\textit{left panel}) and secondary (\textit{right panel}) stars. \textit{Bottom row:} Fourier transform of these lines (in black) and best-match rotational profile (in red) for the primary (\textit{left panel}) and secondary (\textit{right panel}) stars {\it (continued)}. }
\end{figure*}

\begin{figure*}[htb]
\ContinuedFloat
\setcounter{subfigure}{4}
\begin{center}
\subfigure[He\,{\sc i} $\lambda$\,4713 line.]{\includegraphics[width=\linewidth, trim=0.5cm 7.8cm 0cm 2cm, clip=true]{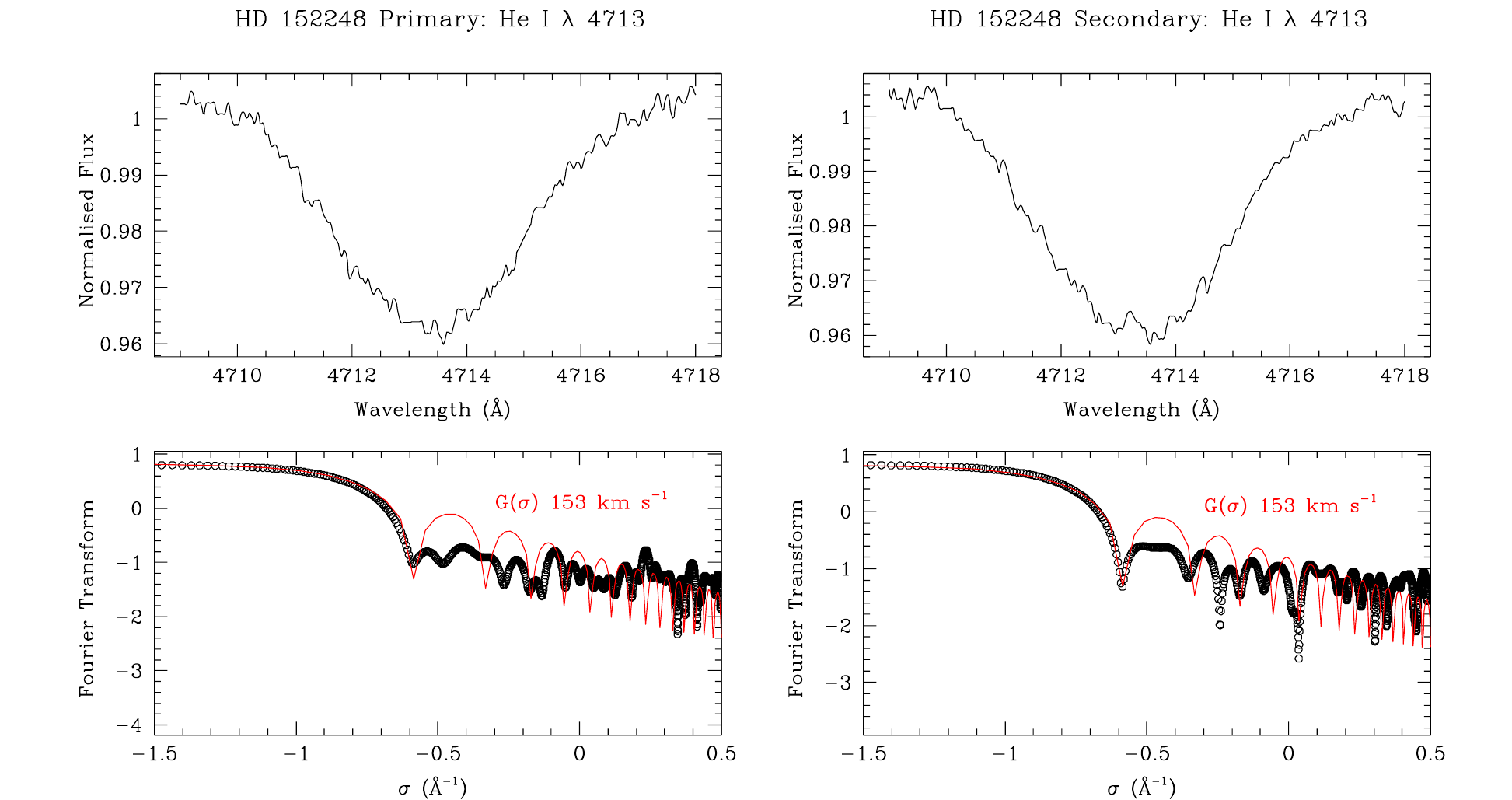}\label{subfig:HeI4713}}
\subfigure[He\,{\sc i} $\lambda$\,4922 line.]{\includegraphics[width=\linewidth, trim=0.5cm 7.8cm 0cm 2cm, clip=true]{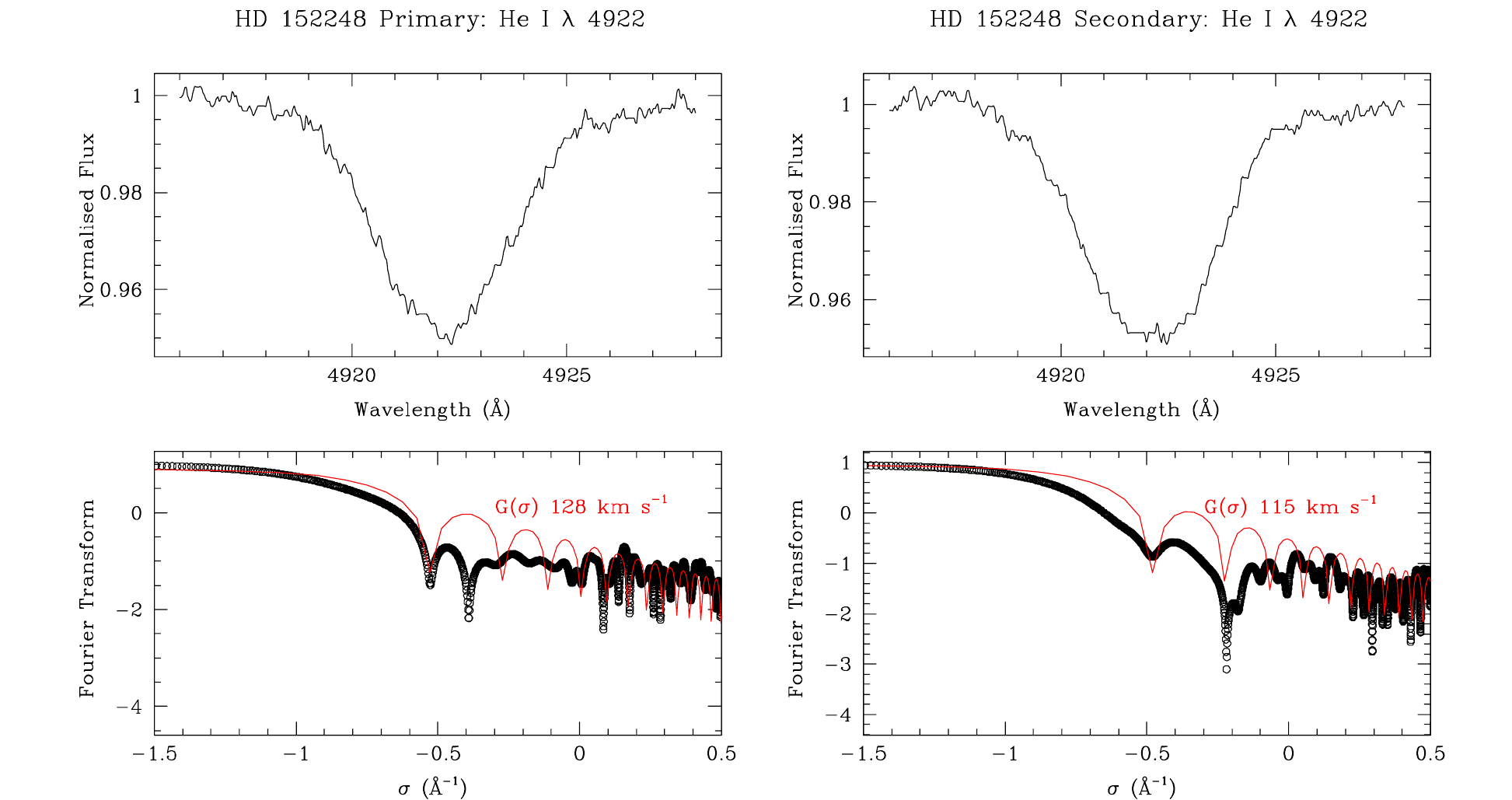}\label{subfig:HeI4922}}
\end{center}
\caption{\textit{Top row:} Line profiles (\subref{subfig:SiIV} Si\,{\sc iv} $\lambda$\,4089, \subref{subfig:OIII} O\,{\sc iii} $\lambda$\,5592, \subref{subfig:CIV} C\,{\sc iv} $\lambda$\,5812, \subref{subfig:HeI4120} He\,{\sc i} $\lambda$\,4120, \subref{subfig:HeI4713} He\,{\sc i} $\lambda$\,4713, \subref{subfig:HeI4922} He\,{\sc i} $\lambda$\,4922, and \subref{subfig:HeI5016} He\,{\sc i} $\lambda$\,5016)  of the separated spectra obtained after application of the brightness ratio for the primary (\textit{left panel}) and secondary (\textit{right panel}) stars. \textit{Bottom row:} Fourier transform of these lines (in black) and best-match rotational profile (in red) for the primary (\textit{left panel}) and secondary (\textit{right panel}) stars {\it (continued)}. }
\end{figure*}

\begin{figure*}[htb]
\ContinuedFloat
\setcounter{subfigure}{6}
\begin{center}
\subfigure[He\,{\sc i} $\lambda$\,5016 line.]{\includegraphics[width=\linewidth, trim=0.5cm 7.8cm 0cm 2cm, clip=true]{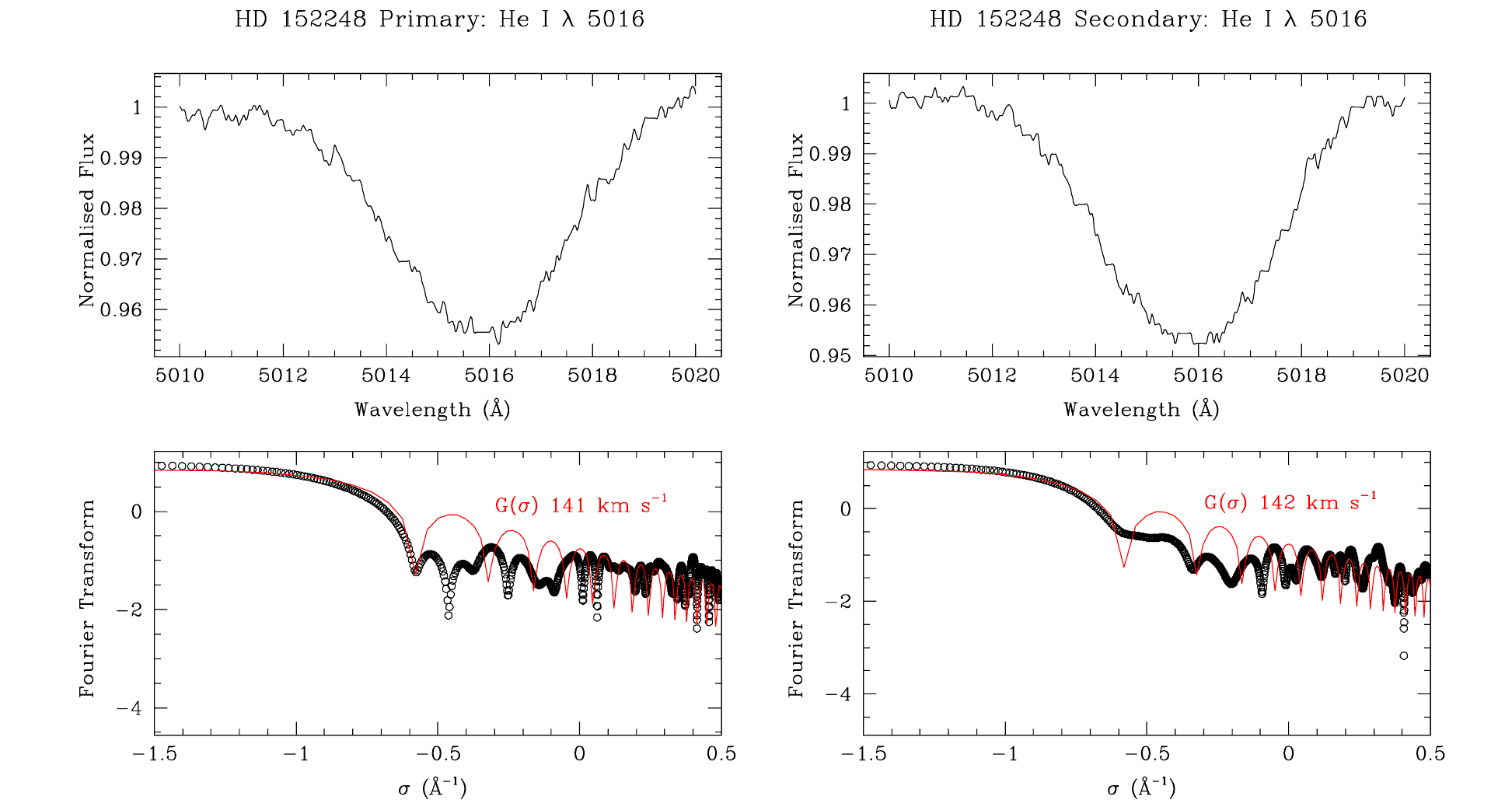}\label{subfig:HeI5016}}
\end{center}
\caption{\textit{Top row:} Line profiles (\subref{subfig:SiIV} Si\,{\sc iv} $\lambda$\,4089, \subref{subfig:OIII} O\,{\sc iii} $\lambda$\,5592, \subref{subfig:CIV} C\,{\sc iv} $\lambda$\,5812, \subref{subfig:HeI4120} He\,{\sc i} $\lambda$\,4120, \subref{subfig:HeI4713} He\,{\sc i} $\lambda$\,4713, \subref{subfig:HeI4922} He\,{\sc i} $\lambda$\,4922, and \subref{subfig:HeI5016} He\,{\sc i} $\lambda$\,5016)  of the separated spectra obtained after application of the brightness ratio for the primary (\textit{left panel}) and secondary (\textit{right panel}) stars. \textit{Bottom row:} Fourier transform of these lines (in black) and best-match rotational profile (in red) for the primary (\textit{left panel}) and secondary (\textit{right panel}) stars {\it (continued)}. }
\end{figure*}

\section{Evidence for a triple system? \label{appendix}}
\citet{Mason98} reported on speckle interferometry data of HD~152\,248 that revealed a nearby object at an angular separation of 0.05\arcsec on two observations out of four. \citet{Mason09} listed a difference of two\,magnitudes compared to the HD~152\,248 binary. These authors also indicated that the detections were uncertain. The mere existence of this object and its physical association with HD~152\,248 are thus unclear and require confirmation. Assuming a physical connection between this object and the eclipsing binary, \citet{Mason98} estimated an orbital period of 150\,yr for an assumed distance of 1.9\,kpc. Scaling this estimate to the {\it Gaia}-DR2 distance yields a putative period of 115\,yr for a circular orbit observed at quadrature.

\begin{figure}[htb]
  \includegraphics[width=0.95\linewidth]{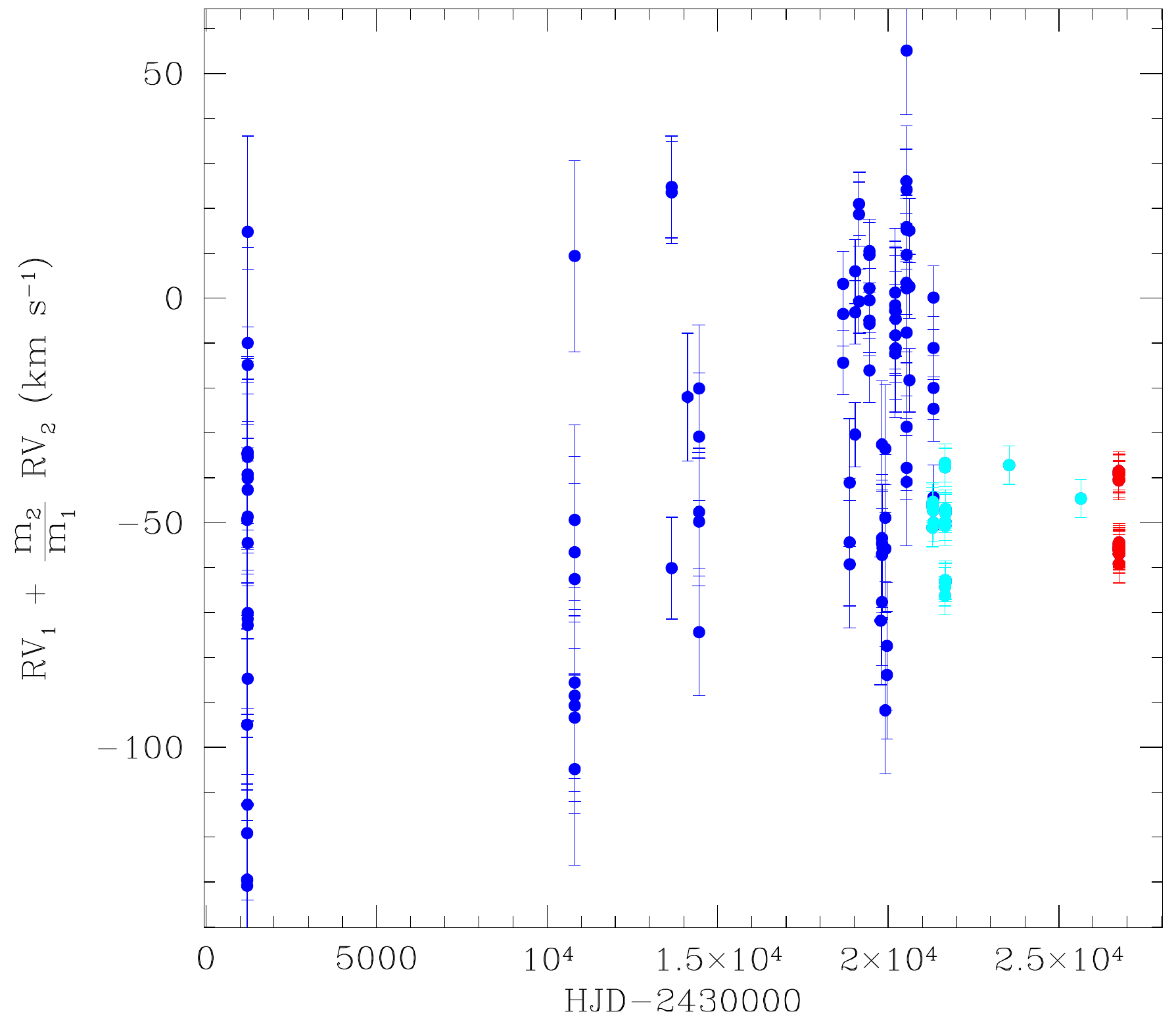}
\caption{Sum of the RV of the primary star and RV of the secondary star times the mass ratio for all spectroscopic observations as computed from Eq.\,\ref{eqn:RVbinary}. The FEROS data are plotted in cyan, the ESPaDOns data in red, and all other data in blue. \label{fig:RVbinary}}
\end{figure}

We reassessed the possible presence of a third component in the system by searching for reflex motion of the binary's centre of mass in the RVs of the binary. We used Eq.\,\ref{massratio}, where we assumed that both stars have the same systemic velocity, meaning $\gamma_{\rm P}  = \gamma_{\rm S} \equiv \gamma$, which gives the expression
\begin{equation}
\label{eqn:RVbinary}
  RV_{\rm P}(t) + \frac{m_2}{m_1}\,RV_{\rm S}(t) = \left(1+\frac{m_2}{m_1}\right)\gamma.
\end{equation}
Assuming $m_2/m_1 = 1.01$, we computed the left-hand side of Eq.\,\ref{eqn:RVbinary} for all the data we have. The result is presented in Fig.\,\ref{fig:RVbinary}.  The oldest data display a huge dispersion, even within short observing campaigns of a few days. This is entirely inconsistent with much slower variations expected for a wide tertiary orbit. The most recent RV values determined through spectral disentangling of the FEROS and ESPaDOns data (cyan and red in Fig.\,\ref{fig:RVbinary}, respectively) display a significantly lower dispersion. Considering these more recent data, we might have the impression of a slow variation, characteristic of a highly eccentric tertiary orbit along with more rapid variations at certain epochs that might then correspond to periastron passages. This would indicate an orbital period of about a decade rather than a century, and would imply a higher multiplicity system with at least four components (including the astrometric companion of \citealt{Mason98}). However, this scenario is not plausible. We observe a large variation between the two sets of ESPaDOns data that were taken during two consecutive nights. Such a fast variation seems inconsistent with a stable hierarchical triple system where the orbital period of the third star should be much longer than that of the inner binary. Therefore, either our estimates of the systemic velocities are biased or the ESPaDOns campaign occurred exactly on the periastron passage of the third component, which seems highly improbable. Focusing on the second FEROS campaign reveals another problem with the tertiary orbit interpretation. During this campaign we observe large variations (see Fig.\,\ref{fig:RVbinary}) that occur on the timescale of the orbital period of the binary, even though these variations should have been removed in the present analysis. We tested different values of the mass ratio $m_2/m_1$ ranging from 0.99 to 1.05 and found that its actual value has an effect on the dispersion. However, none of the values in this range allowed us to remove the 5.816-day periodic variations. These variations most probably reflect the small shape differences between the RV curves of the primary and secondary stars, which can be seen in Table\,\ref{bestfitTable} as the differences in $e$ and $\omega_0$. We suggest that these differences in shape of the RV curve stem from the various interactions between the stars, such as mutual heating and dynamical tides in an eccentric binary \citep[see, e.g.,][]{Moreno05,Palate13}, as well as from wind-wind interactions. This renders the search for a third component in the system by means of the RVs impossible.

We conclude that the existing spectroscopic and photometric data reveal no indication of a triple system: no tertiary lines are apparent in the spectra, and the light curve fits did not require addition of a third light component.  However, we might not be able to confirm the presence or absence of a third component in HD~152\,248 simply based on the analysis of the RVs of the binary system.

\end{appendix}

\end{document}